\documentclass[namedreferences]{SolarPhysics}
\usepackage[optionalrh,natbib]{spr-sola-addons}
\usepackage{graphicx}
\usepackage{url}

\def\km{\;\mbox{km}}

\newcommand{\be}{\begin{equation}}
\newcommand{\ee}{\end{equation}}
\newcommand{\beq}{\begin{eqnarray}}
\newcommand{\eeq}{\end{eqnarray}}

\begin{document}
\begin{opening}

\title{Constructing semi-empirical sunspot models for helioseismology}
\author{R.~\surname{Cameron} \sep L.~\surname{Gizon}$^*$
   \sep H.~\surname{Schunker} \sep A.~\surname{Pietarila}}
\runningauthor{Cameron et al.}
\runningtitle{Constructing semi-empirical sunspot models for helioseismology}
\institute{Max-Planck-Institut f\"ur Sonnensystemforschung,
                Max-Planck-Stra{\ss}e 2, 37191 Katlenburg-Lindau,  Germany\\
           $*$ Corresponding author email: \url{gizon@mps.mpg.de}}

\date{Received ; accepted }

\begin{abstract}
One goal of helioseismology is to determine the subsurface structure of sunspots.
In order to do so, it is important to understand first the near-surface effects of
sunspots on solar waves, which are dominant. Here we construct simplified,
cylindrically-symmetric  sunspot models, which are designed to capture the magnetic and thermodynamics effects
coming from about 500~km below the quiet-Sun $\tau_{5000}=1$ level to the
lower chromosphere.
We use a combination of existing semi-empirical
models of sunspot thermodynamic structure (density, temperature, pressure): the umbral
model of Maltby et al. (1986) and the penumbral model of Ding and Fang (1989).
The OPAL equation of state tables are used to derive the sound speed profile.
We smoothly merge the near-surface properties
to the quiet-Sun values about 1~Mm below the surface. The umbral and penumbral radii are
free parameters. The magnetic field is added to the
thermodynamic structure, without requiring magnetostatic equilibrium. The vertical
component of the magnetic field is assumed to have a Gaussian horizontal profile, with a
maximum surface field strength fixed by surface observations. The full magnetic
field vector is solenoidal and determined by the on-axis vertical field, which, at the surface,
is chosen such that the field inclination is 45$^\circ$ at the umbral-penumbral boundary.
We construct a particular sunspot model based on SOHO/MDI observations of the sunspot in active
region NOAA 9787. The helioseismic signature of the model sunspot is studied using numerical
simulations of the propagation of f, p$_1$, and p$_2$ wave packets. These simulations are
compared against cross-covariances of the observed wave field. We find that the sunspot model
gives a helioseismic signature that is similar to the observations.
\end{abstract}

\keywords{Sun: Helioseismology, Sun: sunspots, Sun: magnetic fields}

\end{opening}

\section{Introduction}
The subsurface structure of sunspots is poorly known.
Previous attempts to use helioseismology to determine the
subsurface properties of sunspots have mainly been done
under the assumptions that the sunspot is non-magnetic
(it is usually treated as an equivalent sound-speed perturbation) and that it is a weak perturbation to the quiet-Sun.
Neither of these two assumptions is justifiable.
The helioseismology results have been\,---\,perhaps unsurprisingly\,---\,confusing and contradictory \citep[e.g.,][]{gizon09}.
The effects of near-surface magnetic and structural perturbations on solar waves
are strong is more easily dealt with in numerical simulations.
Examples of such simulations of wave propagation through prescribed sunspot models
include \cite{cameron08, hanasoge08, khomenko08, parchevsky09}.
Necessary for all these attempts are appropriate model sunspots. There are numerous sunspot models
available, some of which have been
used in helioseismic studies. For recent reviews of sunspot models see, e.g., \citet{Solanki2003}, \citet{Thomas2008}, and \citet{Moradi10}.
Various authors, e.g., \citet{khomenko08a}, \citet{Moradi2008}, and \citet{Moradi2009ApJ}, have constructed magnetohydrostatic parametric sunspot models for use in helioseismology.
In a previous paper \citep{cameron08}, we considered a self-similar magnetohydrostatic model.

Although the deep structure of sunspots is of the utmost interest, it is likely to be swamped in the helioseismic observations unless we are able to accurately model and remove the surface effects \citep[see, e.g.,][]{Gizon10}. The aim of this paper is to construct a simple parametric sunspot model, which captures the main effects of the near-surface layers of sunspots on the waves.

In principle, the 3D properties (magnetic field, pressure, density, temperature, Wilson depression, flows) of sunspots can be inferred in the photosphere and above using spectropolarimetric inversions \citep[e.g.,][]{mathew03}.
However, today, these inversions are unavailable for most sunspots.
In such circumstances, we choose to construct sunspot models that are based upon
semi-empirical models of the vertical structure of typical sunspots.
In producing our models, we treat separately the thermodynamic structure and the magnetic field:
 we do not require magnetostatic equilibrium. There are several reasons for
not requiring the model to be hydrostatic. The first is that we are
much more interested in the waves then in whether the background model
is magnetodydrostatic. For this reason we are more interested in geting
for example, the sound speed, density, and magnetic field to be close to
those which are observed.
A second reason is that we are not including non-axisymettric structure
in our model, which certainly affects the force-balance.
A third reason is that sunspots
have both Evershed and moat flows, the existence of which implies
a net force and so indicates that the sunspot is not strictly
magnetohydrostatic.

In this paper, we take the umbral and penumbral models
to match those of existing semi-empirical models.
In the absence of horizontal magnetic field measurements we assume that the
field inclination at the umbra/penumbra boundary is approximately $45^{\circ}$.
We do not consider the effects of the Evershed or moat flows,
although they could be included in the framework we set out.
The surface model of the sunspot needs to be smoothly connected to the quiet-Sun
model below the surface, and details will be provided in Section 2.
Since we are not including the chromosphere in our simulations, we have chosen
to smoothly transition the sunspot model above the surface to the quiet-Sun model.

The description for constructing models including the surface structure
of sunspots, which will be fleshed out in Section 2, is intended to be
generic. For illustrative purposes we choose parameter values of the sunspot model that are appropriate for the
sunspot in NOAA 9787, which was observed by SOHO/MDI in January 2002 \citep{gizon09}.
In Section~3, we describe the setup of the numerical simulation of the propagation of f, p$_1$, and p$_2$ wave packets through the model sunspot. We use the SLiM code \citep{cameron07}.
We briefly compare the simulations and the SOHO/MDI observations in Section~4.
In Section~5, we conclude that this simple sunspot model, which is intended to
be a good description of the sunspot's surface properties, leads to a good
 helioseismic signature and provides a testbed for future studies.


\section{Model sunspot}
The sunspot models that we describe below are cylindrically symmetric and thus we use a
cylindrical-polar coordinate system to describe the spot, where $z$ is the height and $r$ is the horizontal
distance from the axis of the sunspot.

\subsection{Thermodynamical aspects}
For this paper we use the umbral Model-E  of \cite{maltby86} and
the penumbral model of \cite{ding89}.
These models give the pressure and density as functions of height.
For the quiet-Sun background in which we embed the sunspot, we use Model S \citep{JCD86}.
Zero height ($z=0$) is defined as in Model S.

\begin{figure}
\includegraphics[width=0.45\textwidth]{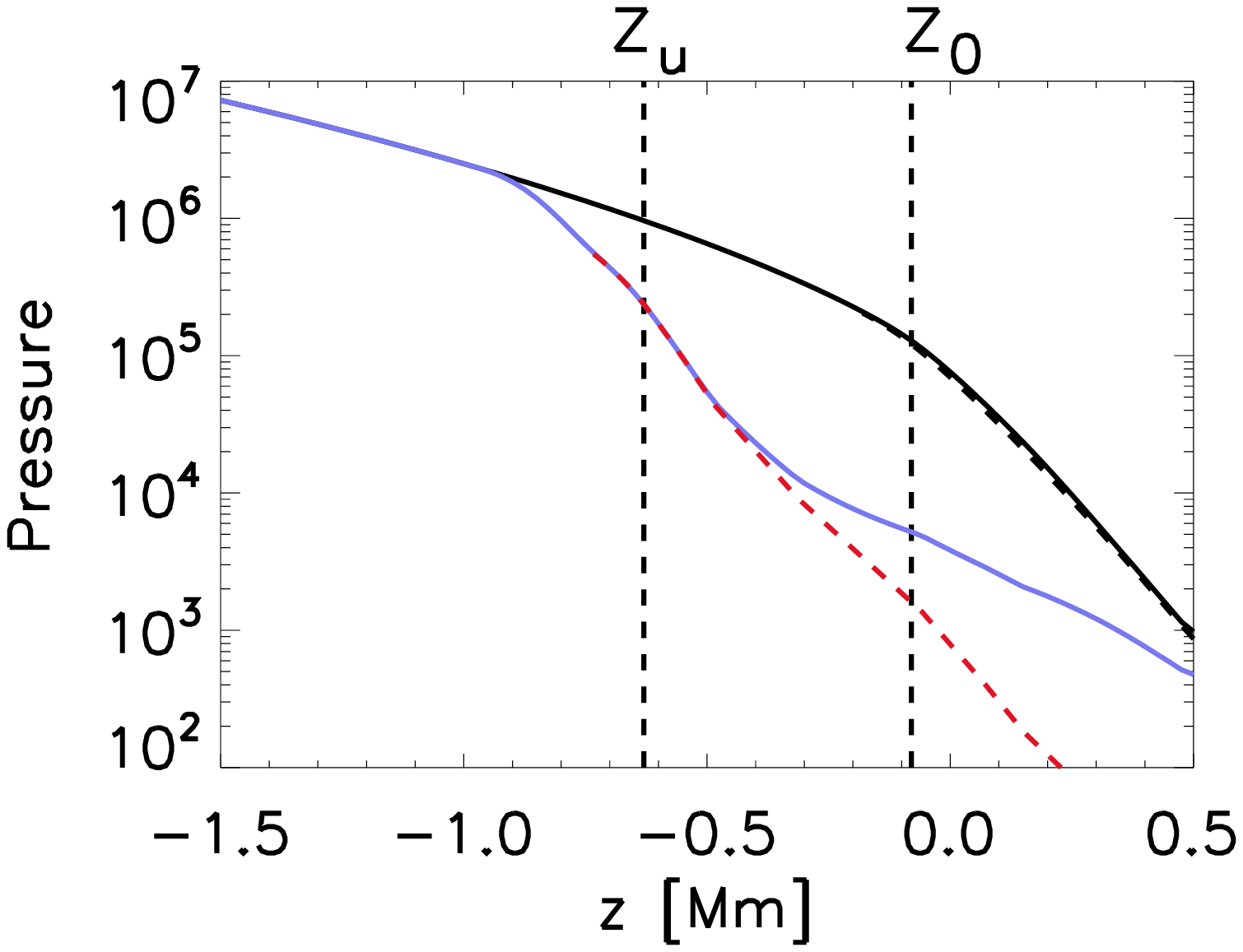}
\includegraphics[width=0.45\textwidth]{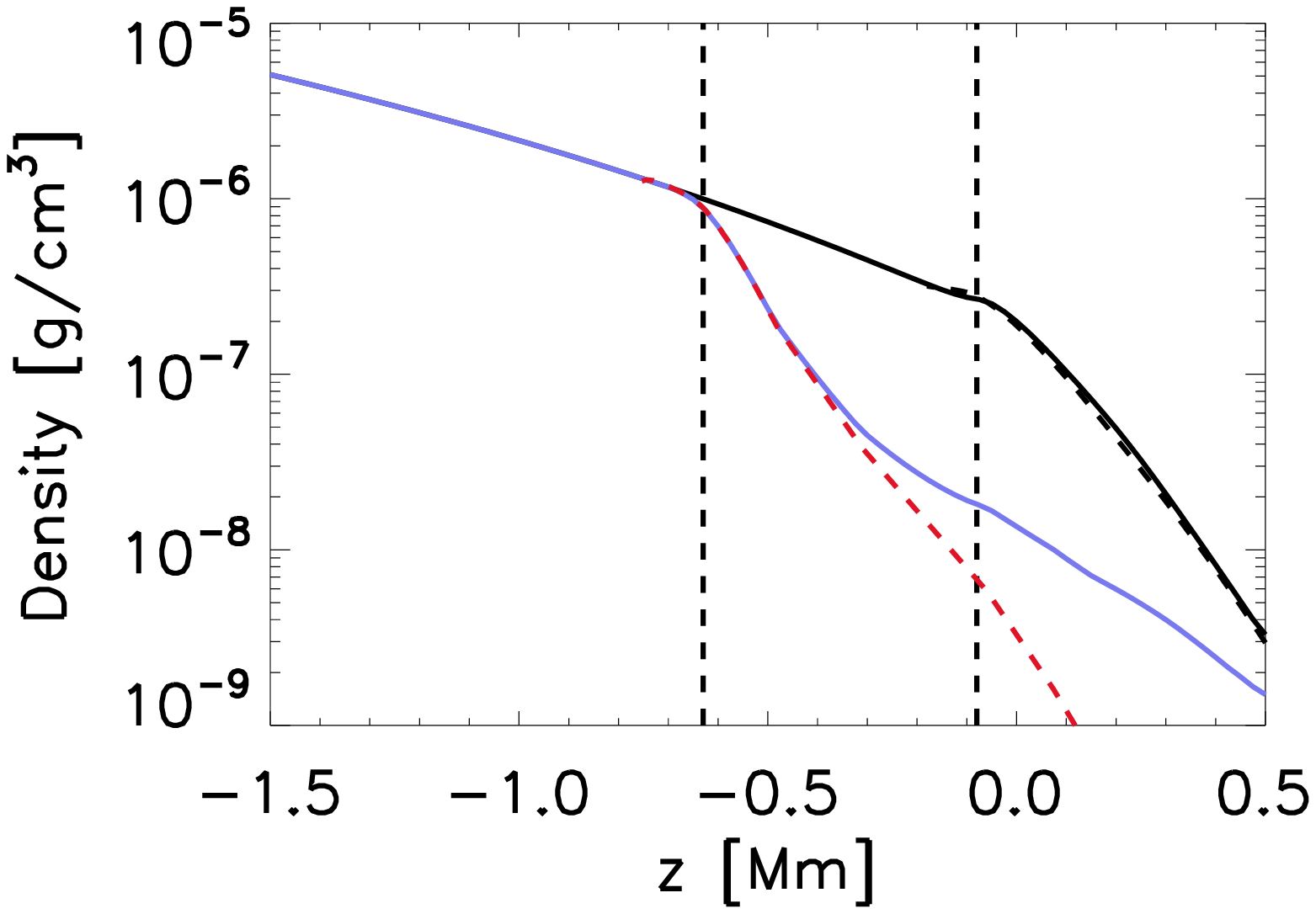}

\includegraphics[width=0.45\textwidth]{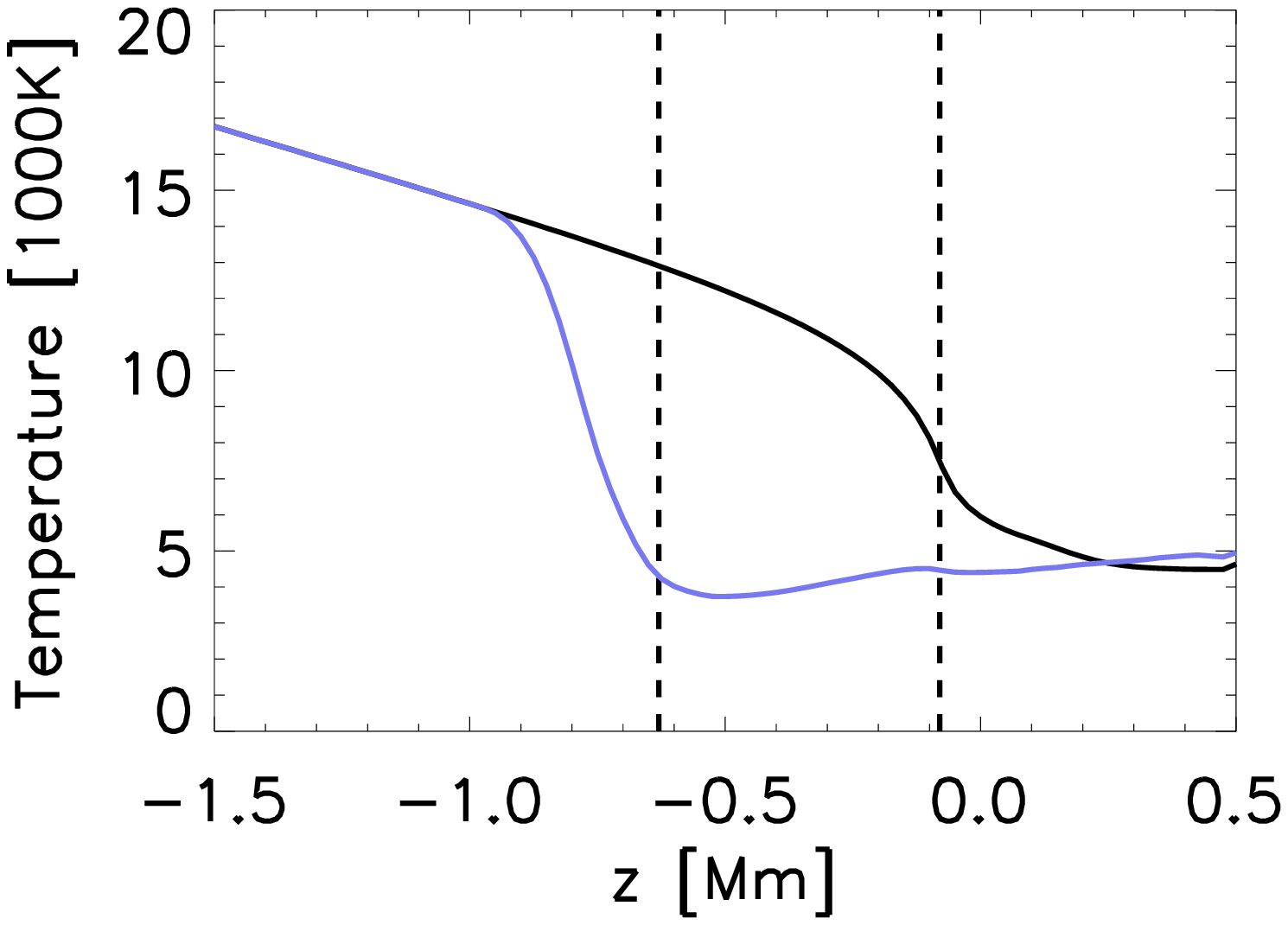}
\includegraphics[width=0.45\textwidth]{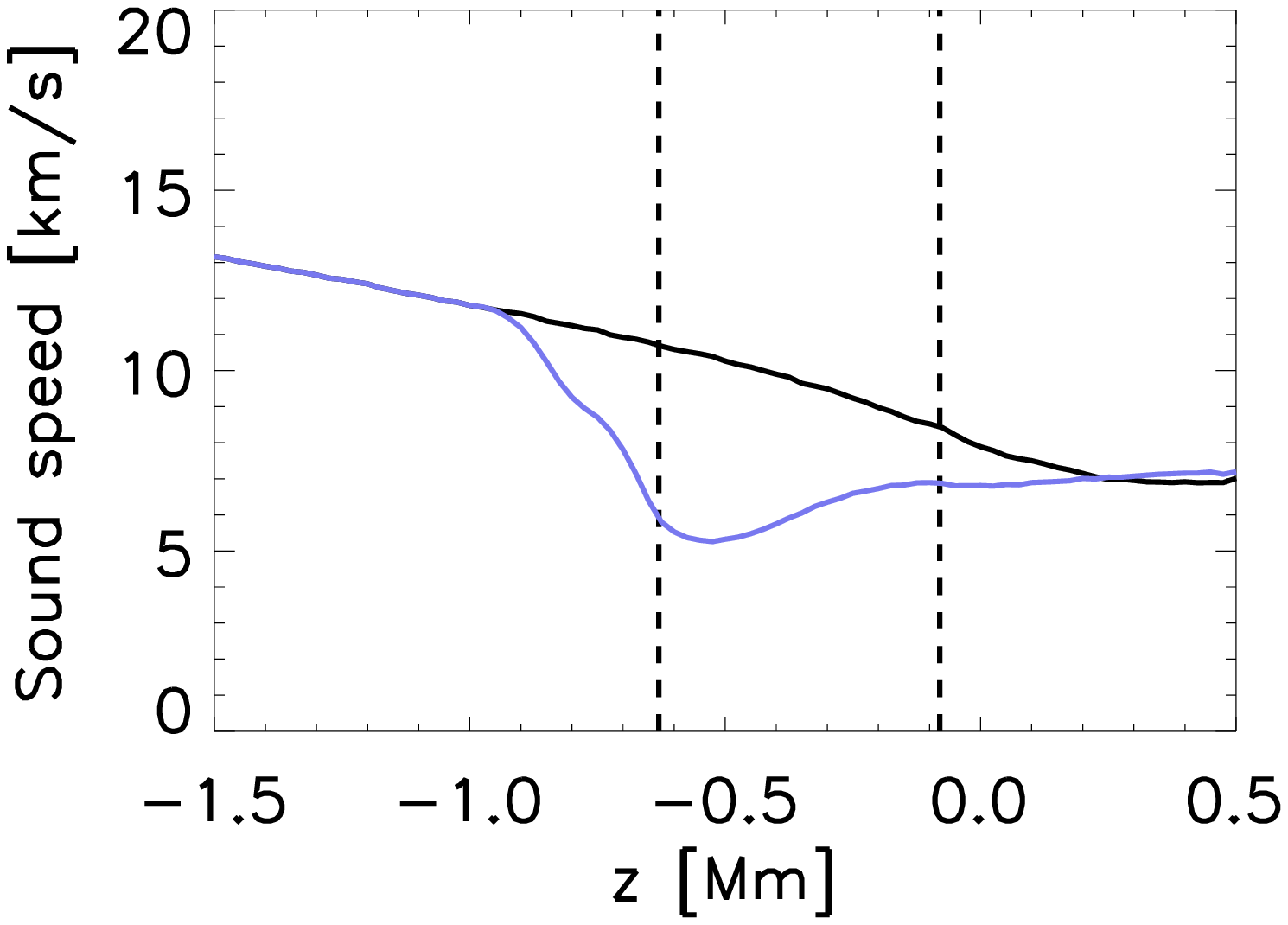}

\caption{The red-dashed curves show the semi-empirical umbral model from \cite{maltby86} as a function of height.
The black curves are the quiet-Sun values from model-S. The blue curves show the pressure the umbral model described in this paper.
The panels show: presssure (top-left panel), density (top-right),
temperature (bottom-left), and sound-speed (bottom-right).
The vertical dashed lines are the heights where $\tau_{5000}=1$
in the quiet-Sun ($z=z_0$) and the umbra ($z=z_{\rm u}$).}
\label{fig_atm_umbra}
\end{figure}

There is some freedom in choosing the depths of the umbral (Wilson) and penumbral depressions.
We choose a Wilson depression of 550~km.   
More precisely, we place the $\tau_{5000}=1$ surface of the umbra at a height
\begin{equation}
z_{\rm{u}} = z_0 - 550 \km,
\end{equation}
which is defined with respect to the height of the $\tau_{5000}$ height of the quiet Sun reference model of  \cite{maltby86}
at with $z_0$ being approximately $-70$ km. This value of $550$~km for the Wilson depression produces a match between the density of the Maltby
model and that of the quiet Sun approximately 100~km below the $\tau_{5000}=1$ surface (see Figure~\ref{fig_atm_umbra}).

For the penumbra, we place the $\tau_{5000}=1$ surface at a height
\begin{equation}
z_{\rm{p}} = z_0 - 150  \km.
\end{equation}
A penumbral depression of 150~km is consistent with spectropolarimetric measurements \citep[e.g.,][]{Mathew2004}.

The umbral model of \citet{maltby86} is plotted in Figure~\ref{fig_atm_umbra} for $z>z_{\rm u} - 116\; \rm{km}  = -736$~km.
The penumbral model of \citet{ding89} is plotted in Figure~\ref{fig_atm_umbra} for $z>z_{\rm p} - 220\; \rm{km} = -440$~km.

\begin{figure}
\includegraphics[width=0.45\textwidth]{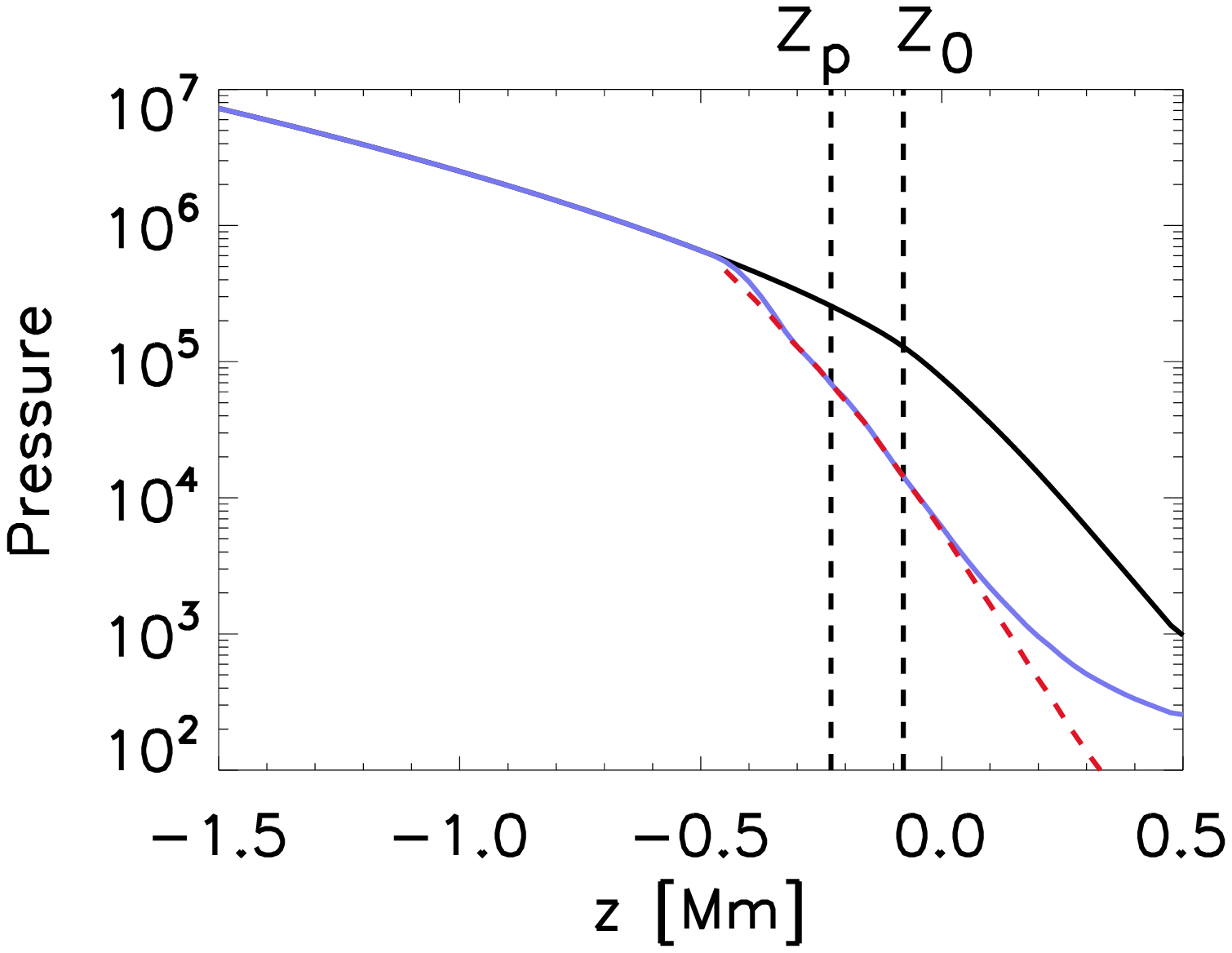}
\includegraphics[width=0.45\textwidth]{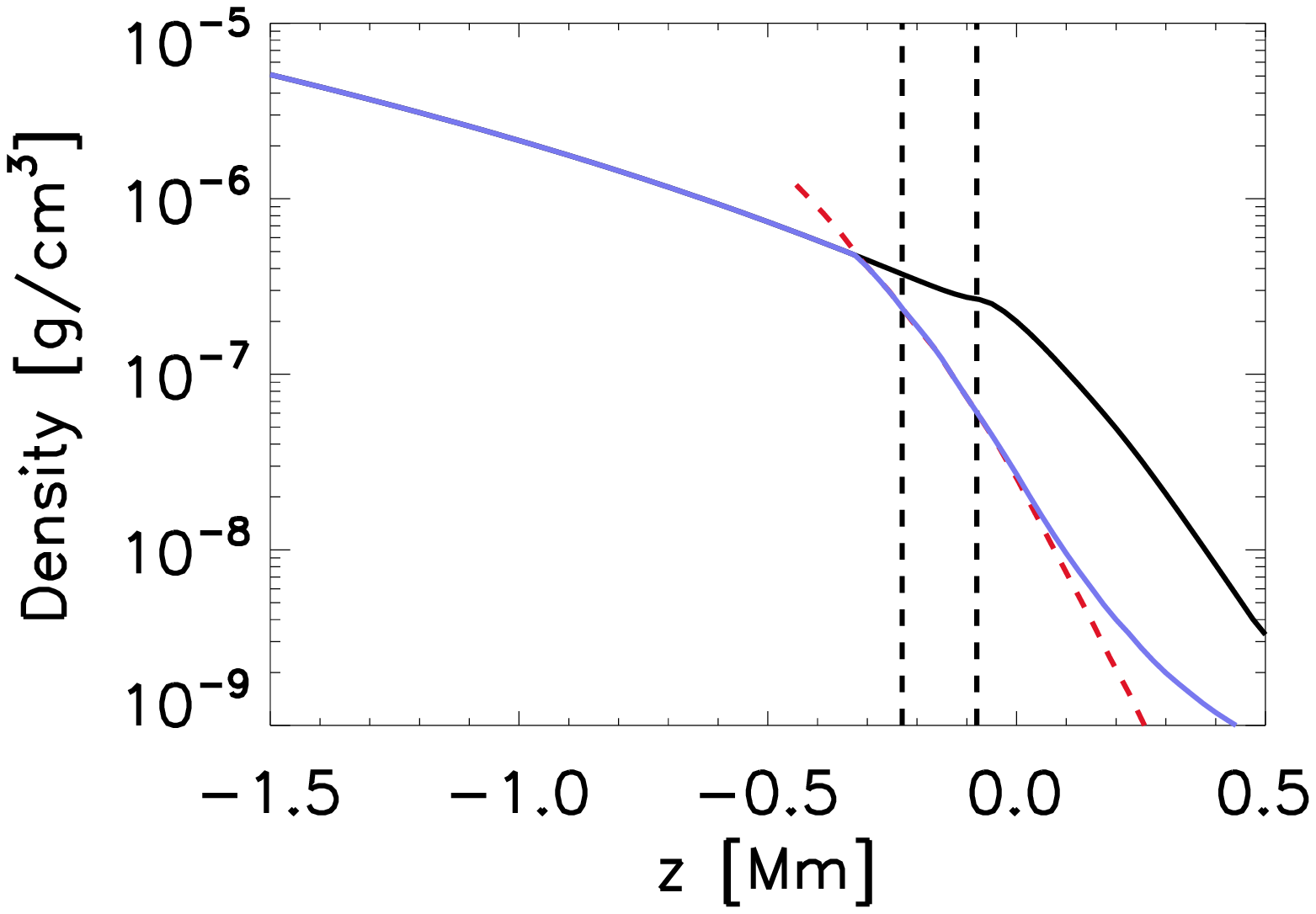}

\includegraphics[width=0.45\textwidth]{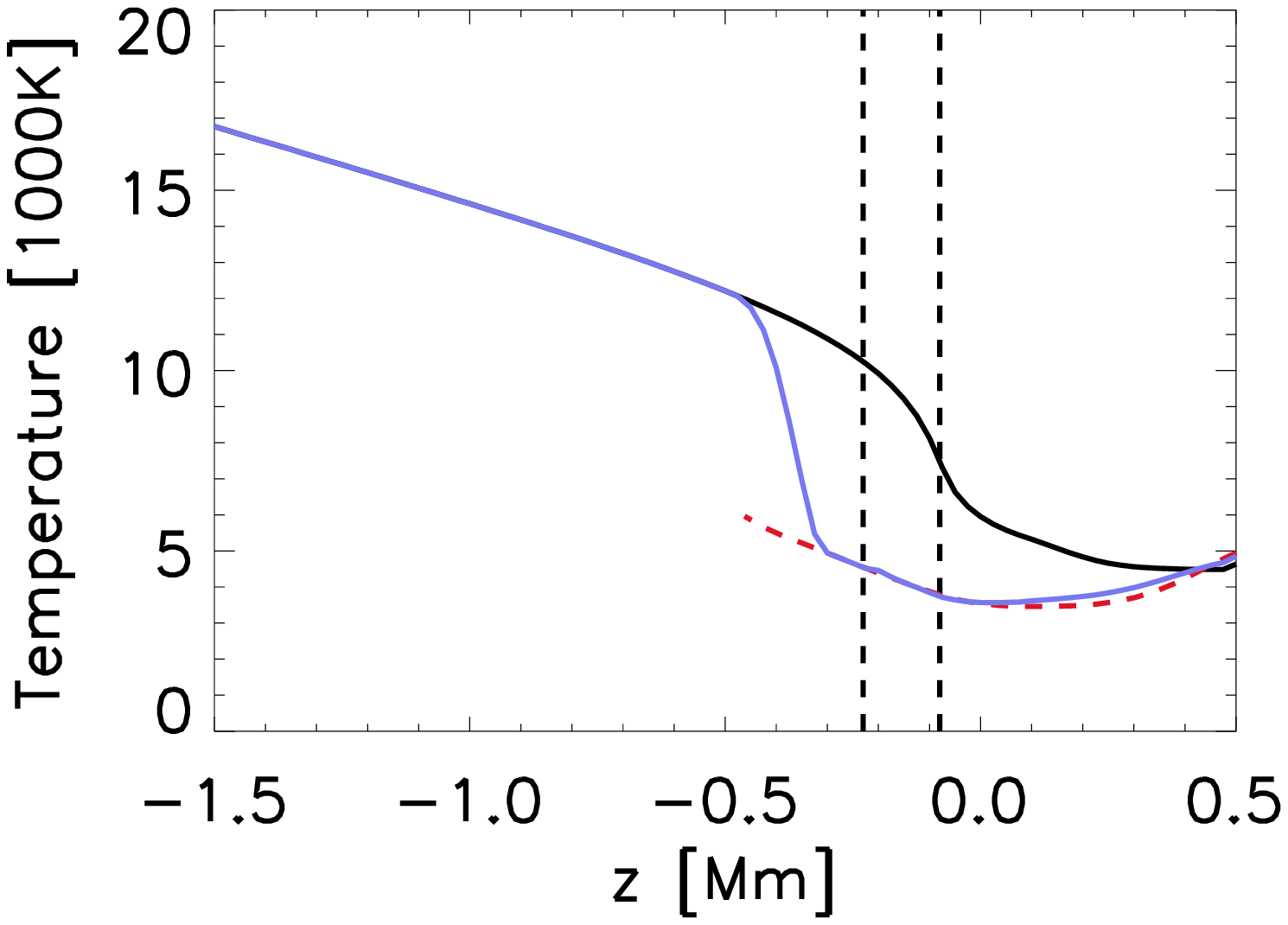}
\includegraphics[width=0.45\textwidth]{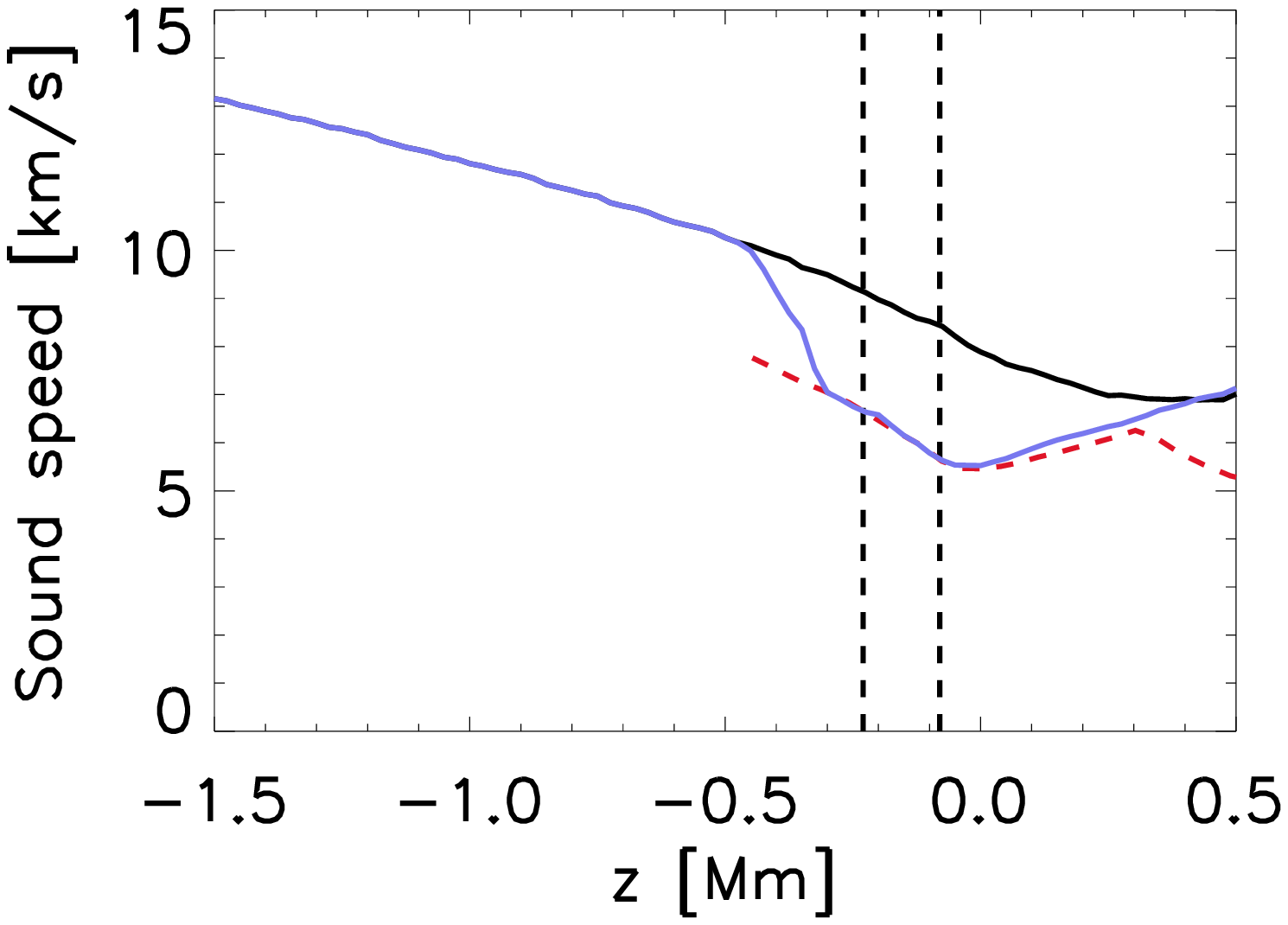}

\caption{The red-dashed curves show the semi-empirical penumbral model from \cite{ding89}.
The black curves are the quiet-Sun values from Model S.
The blue curves show the penumbral model described in this paper.
Shown are, the pressure (top-left panel), density (top-right),
temperature (bottom-left), and sound-speed (bottom-right).
The vertical dashed lines are the heights where $\tau_{5000}=1$
in the quiet-Sun ($z=z_0$) and the penumbra ($z=z_{\rm p}$).}
\label{fig_atm_penumbra}
\end{figure}

\subsection{Geometrical aspects}
The umbral and penumbral models discussed above need to be smoothly embedded in Model S.
In order to do so, we modify the semi-empirical models above and below the $\tau_{5000}=1$ levels of the umbra and the penumbra, as follows.

In our model, the pressure along the axis of the sunspot, denoted by $P_{\rm u}$, is such that
\begin{equation}
\ln P_{\rm u}(z) = w(z) \ln P_{\mathrm{Maltby}}(z)  + (1-w(z))  \ln P_{\mathrm{qs}}(z) .
\end{equation}
In this expression, $P_{\mathrm{Maltby}}(z)$ is the umbral pressure given by \citet{maltby86} for $z>-736$~km, and is extrapolated for $z<-736$~km assuming constant pressure scale height. The quantity $P_{\mathrm{qs}}(z)$ is the quiet-Sun pressure from Model S.
The function $w(z)$ is a weighting function given by
\begin{equation}
w(z)=    \left\{
\begin{array}{l l}
0& z  \le z_{\rm{u}} -380\km ,\\
\cos[\pi \frac{(z_{\rm{u}} - 80 \km - z)}{300 \km}] /2  + 1/2 &z_{\rm{u}} - 380 \km < z \le z_{\rm{u}} - 80 \km ,\\
 1   & z_{\rm{u}} - 80 \km  < z \le  z_{\rm{u}} + 20 \km ,\\
\cos[\pi \frac{(z_{\rm{u}} + 20 \km - z)}{(900 \km - z_{\rm{u}})}]/2 + 1/2 &   z_{\rm{u}} + 20 \km  < z \le 900 \km , \\
0 & 900 \km<z .
\end{array} \right.
\end{equation}
The resulting umbral pressure is plotted in Figure~\ref{fig_atm_umbra}.  The pressure near $z=z_{\rm u}$ is that of \citet{maltby86}. Below, it smoothly merges with the quiet Sun pressure at $z=z_{\rm u} - 380 \; \rm{km} = - 1000$~km. Above, the pressure tends to the quiet-Sun value and there is a significant departure from \citet{maltby86} for $z \gtrsim -200$~km  (by more than a factor of two).

The vertical profile of the density is treated in the same manner as the pressure. The Maltby model does not explicitly contain all the properties that we require, so we use the OPAL tables to derive the sound speed from the pressure and the density.

The penumbral pressure and the density of \citet{ding89} are embedded in Model S in a similar way as was done
for the umbra, except that the weighting function $w(z)$ is replaced by
\begin{equation}
w(z)=    \left\{\begin{array}{l l}
0& z  \le z_{\rm{p}} -280\km,\\
\cos[\pi \frac{(z_{\rm{p}} - 80 \km - z)}{200 \km}] /2  + 1/2 &z_{\rm{p}} - 280 \km < z \le z_{\rm{p}} - 80 \km ,\\
 1   & z_{\rm{p}} - 80 \km  < z \le  z_{\rm{p}} + 20 \km ,\\
\cos[\pi \frac{(z_{\rm{p}} + 20 \km - z)}{(800 \km - z_{\rm{p}})}]/2 + 1/2 &   z_{\rm{p}} + 20 \km  < z \le 800 \km\\
0 & 800 \km < z.
\end{array} \right.
\end{equation}
The resulting penumbral pressure and density are plotted in Figure~\ref{fig_atm_penumbra}.

Thus far we have described umbral, penumbral, and quiet-Sun models.
We use them to form a three-dimensional cylindrically-symmetric sunspot model.

Since we are modeling only the very near-surface layers of the sunspot (top ~1 Mm), we do not
pay excessive attention to the fanning of the interfaces with depth.
In our 3D sunspot model, each thermodynamic quantity is the product of a function of $z$ and a
function of $r$.

\begin{figure}
\includegraphics[width=0.45\textwidth]{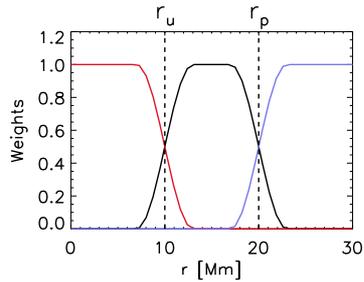}
\caption{Relative weights used to combine the umbral (red), penumbral (black), and quiet-Sun (blue) models in the horizontal radial direction.}
\label{fig_weights}
\end{figure}

The radii of the umbra and the penumbra are denoted by $r_{\rm u}$ and $r_{\rm p}$.
We combine the three model components (umbra, penumbra, quiet-Sun) in the $r$ coordinate
using the weight functions shown in Figure~\ref{fig_weights}.
The transitions between the 1-D atmospheres have a width of $6$~Mm and are described by raised cosines.

For the sunspot in Active Region 9787, we take $r_{\rm{u}}=10$~Mm and $r_{\rm{p}}=20$~Mm for the umbral and penumbral radii.

The sound speed was again reconstructed using the density, pressure, and the OPAL tables to ensure consistency.
We note that the temperature does not appear in the equations and is used only for the purpose of calculating the properties of the spectral lines used in helioseismology (e.g., the MDI Ni 676.8~nm line).

\subsection{Observables}

Doppler velocity is the primary observable in helioseismology.
In this section, which is parenthetical, we consider the
question of how to compute a quantity resembling the observations from the sunspot model.
This topic has previously been considered by, e.g.,  \cite{wachter08}.
Here we use the STOPRO code \citep{solanki87} to synthesize, assuming LTE, the formation
of the Ni 676.8~nm line, which is the line used by the SOHO/MDI instrument.

\begin{figure}
\includegraphics[width=0.48\textwidth]{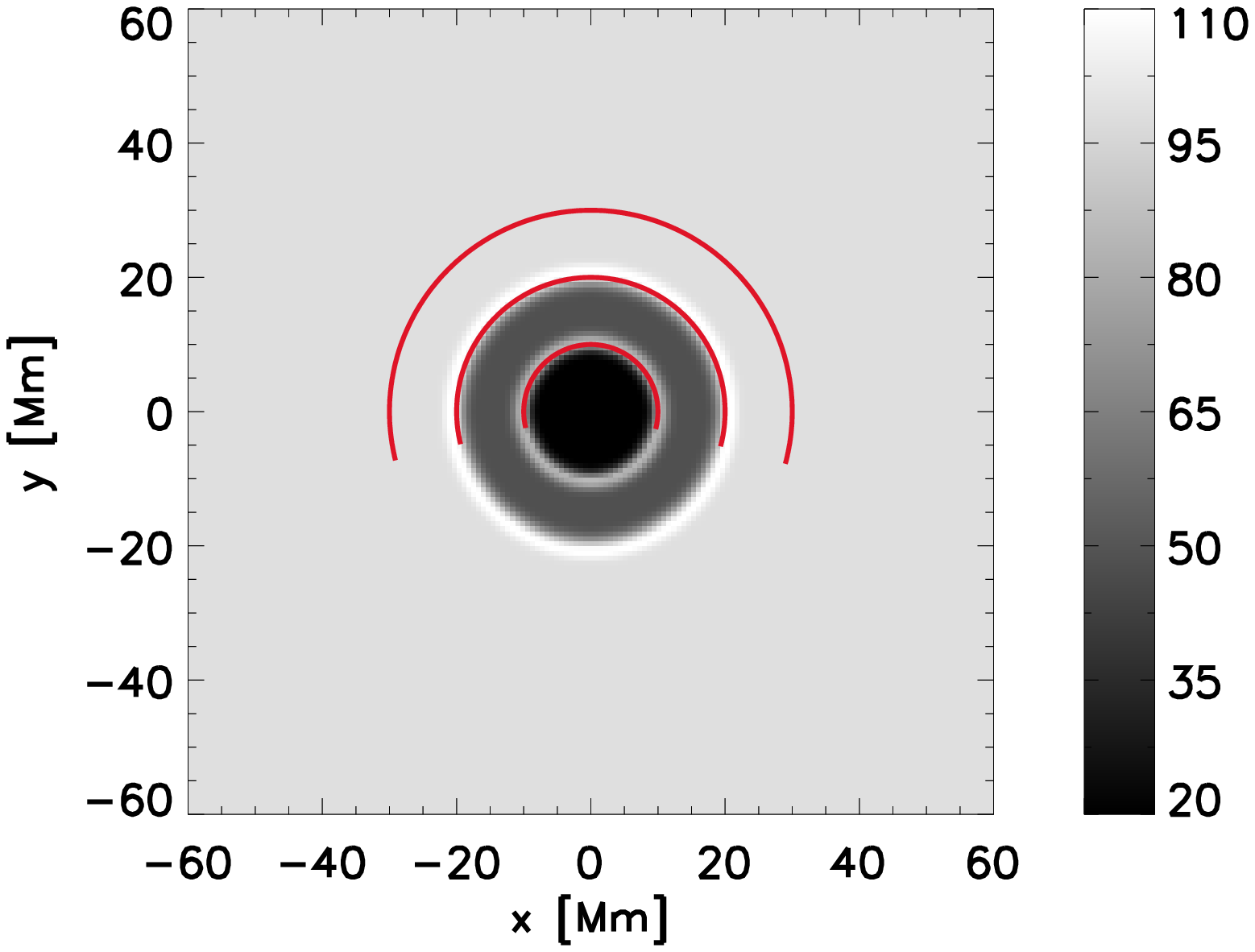}
\includegraphics[width=0.48\textwidth]{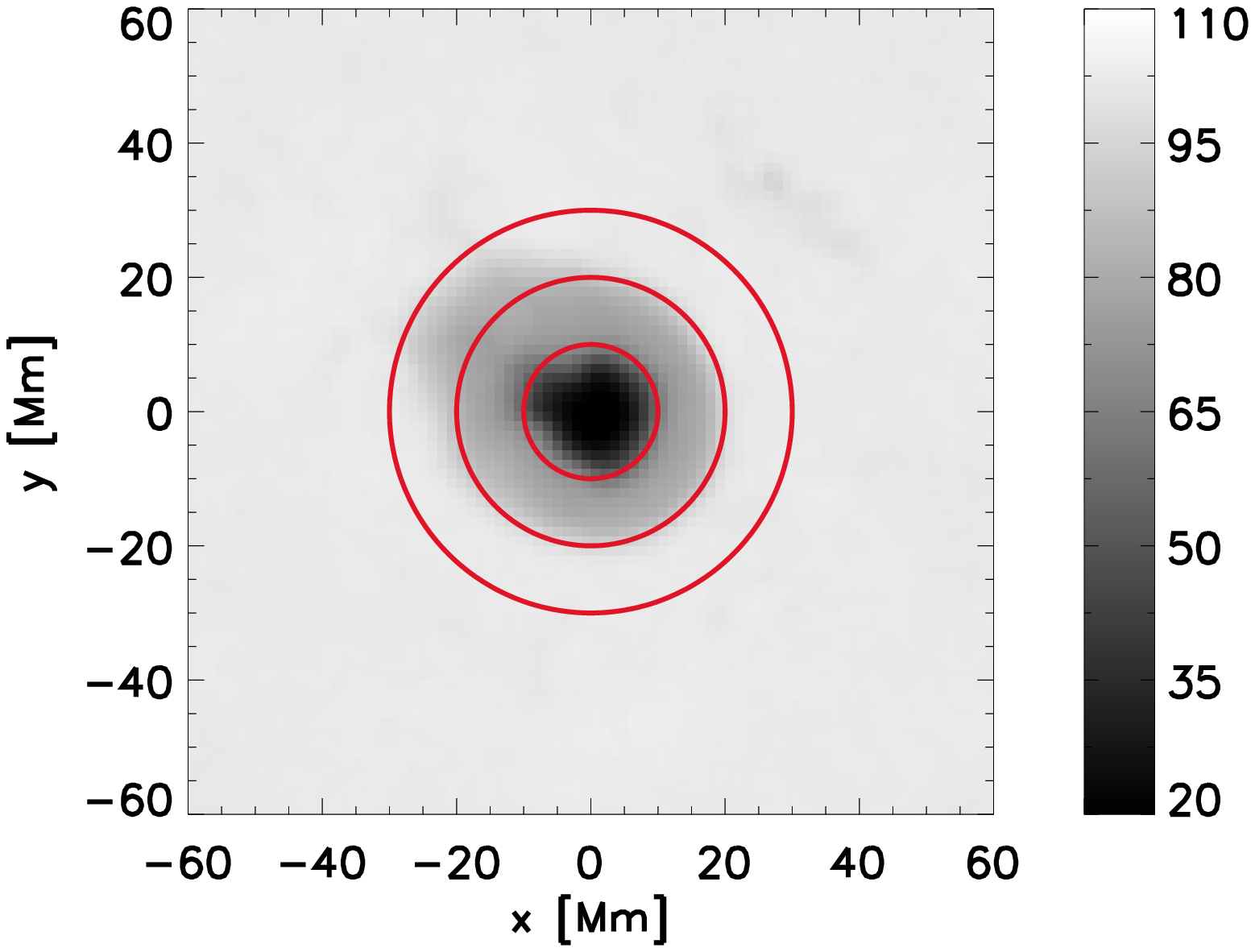}
\caption{Left panel: Model of the sunspot continuum intensity image near the Ni $676.8$~nm line calculated using STOPRO. The red arcs have radii 10 Mm, 20 Mm, and 30 Mm.
Noticeable are bright rings at the transitions between the umbra, penumbra, and quiet Sun.
Right panel: Observed SOHO/MDI intensity image of the sunspot in AR 9787 averaged over 20\,--\,28 January 2002. }
\label{fig_sim_I}
\end{figure}

We obtain a continuum intensity image near the Ni line, shown in  Figure~\ref{fig_sim_I}, together with observations.  The bright rings at the transitions between the umbra, penumbra, and quiet Sun are undesirable artifacts due to the simplified transitions between the different model components.
This may have to be addressed in a future study by adjusting the weights from Figure~\ref{fig_weights}.

\begin{figure}
\includegraphics[width=0.5\textwidth]{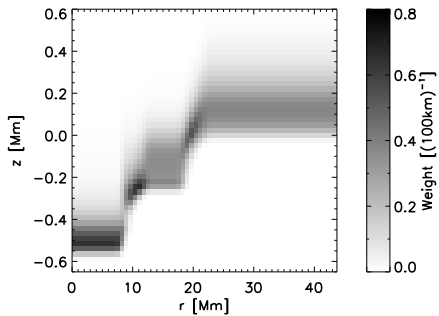}
\caption{The weights (being the normalized response function of the Ni 676.8~nm line)
which allow us to create synthetic helioseismic Doppler velocity maps. The height
is with reference to the $\tau_{5000}=1$ surface in the quiet-Sun. }
\label{fig_sim_rf}
\end{figure}

We compute the response functions for vertical velocity perturbations  \citep{beckers75} as functions of height, horizontal position, and wavelength, $\lambda$. We integrate the response function over the wavelength
bandpass [$-100$~m\AA, $-34$~m\AA] measured from line center.  This wavelength range was chosen because
it is where the slope of the line is maximal, i.e. where the response to velocity perturbations
is largest.  The response function is normalized so that its $z$-integral is unity at each horizontal position. These are the weights (Figure~\ref{fig_sim_rf}) to be multiplied by the vertical velocity and then integrated in the vertical direction to give a Doppler velocity map.

Whether the bright rings affect the Doppler velocity in practice is unclear. We believe, however, that the response function computed here will be useful in studies investigating the diagnostic potential of the helioseismic signal within the sunspot.

\subsection{Magnetic field}
When observations of the vector magnetic field are available, they can and should be
included in constraining the sunspot model. In this paper, we consider a more generic case and assume a Gaussian dependence of the vertical
 magnetic field in the radial direction. The important parameters are then the maximum magnetic field at the surface, $B_0$,
the half-width at half maximum (HWHM) of the Gaussian profile at the surface, $h_0$, and the inclination of the magnetic field at the umbral-penumbral boundary.

Explicitly, the dependence of $B_z$ on $r$ and $z$ is given by
\begin{eqnarray}
B_z(r,z)=B_z(r=0,z) \exp[-(\ln 2)  r^2/ h(z)^2]
\label{eq.radspot}
\end{eqnarray}
where $h(z)$ is the half width at half maximum (HWHM) of the Gaussian.
Since the magnetic flux at each height must be  conserved, we have
\begin{equation}
h(z)=h_0 \sqrt{B_0/B_z(r=0,z)} ,
\end{equation}
where $h_0$ is the surface value of $h$.
For the sunspot in Active Region 9787 we have chosen $B_0 = 3$~kG and $h_0=10$~Mm.

The function $B_z(r=0,z)$ is unknown. It is a goal of helioseismology to constrain $B_z$ along the sunspot axis.
Here we choose a two-parameter function as follows:
\begin{eqnarray}
B_z(r=0,z)=B_0 e^{-(z/\alpha) L(-z/\alpha)L(z/\alpha_2)} ,
\label{eq.logistic}
\end{eqnarray}
where $L(z)=1/(1+e^{-z})$ is the logistic function.
Since $d B_z /dz(r=0,z=0)  = - B_0/(4\alpha)$, the parameter $\alpha$ controls the vertical gradient of $B_z$
near the surface. Together with the condition $\nabla\cdot{\bf B}=0$, $d B_z/dz (r=0,z=0)$ determines the full
surface vector magnetic field. We chose $\alpha=1.25$~Mm  so that at $z=0$ the inclination of the magnetic
field is $45^\circ$ at the umbra/penumbra boundary, in agreement with observations. The parameter $\alpha_2$
controls the field strength at depth;  we chose $\alpha_2=18.4$~Mm.

\begin{figure}
\includegraphics[width=0.45\textwidth]{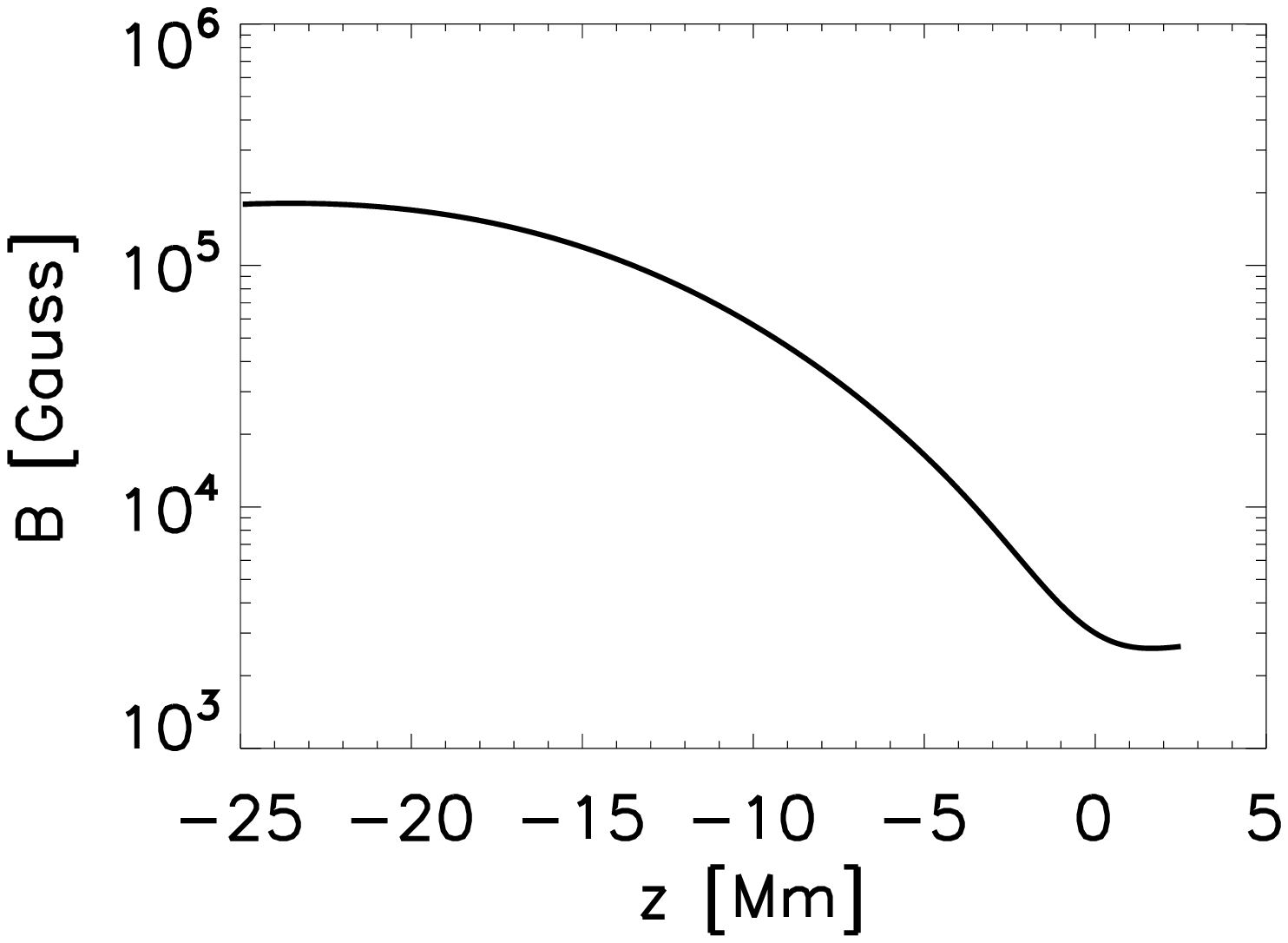}
\includegraphics[width=0.45\textwidth]{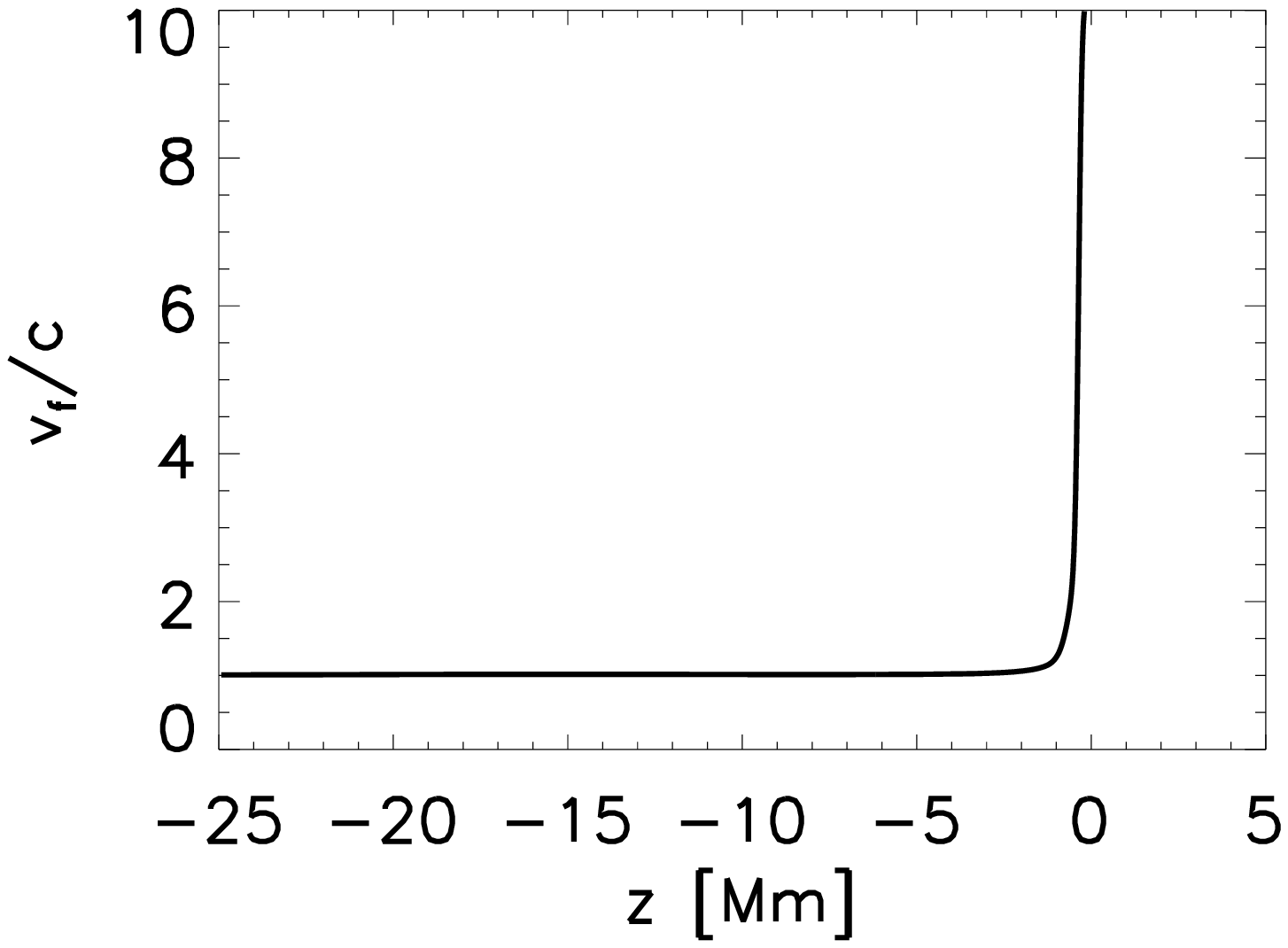}

\includegraphics[width=0.45\textwidth]{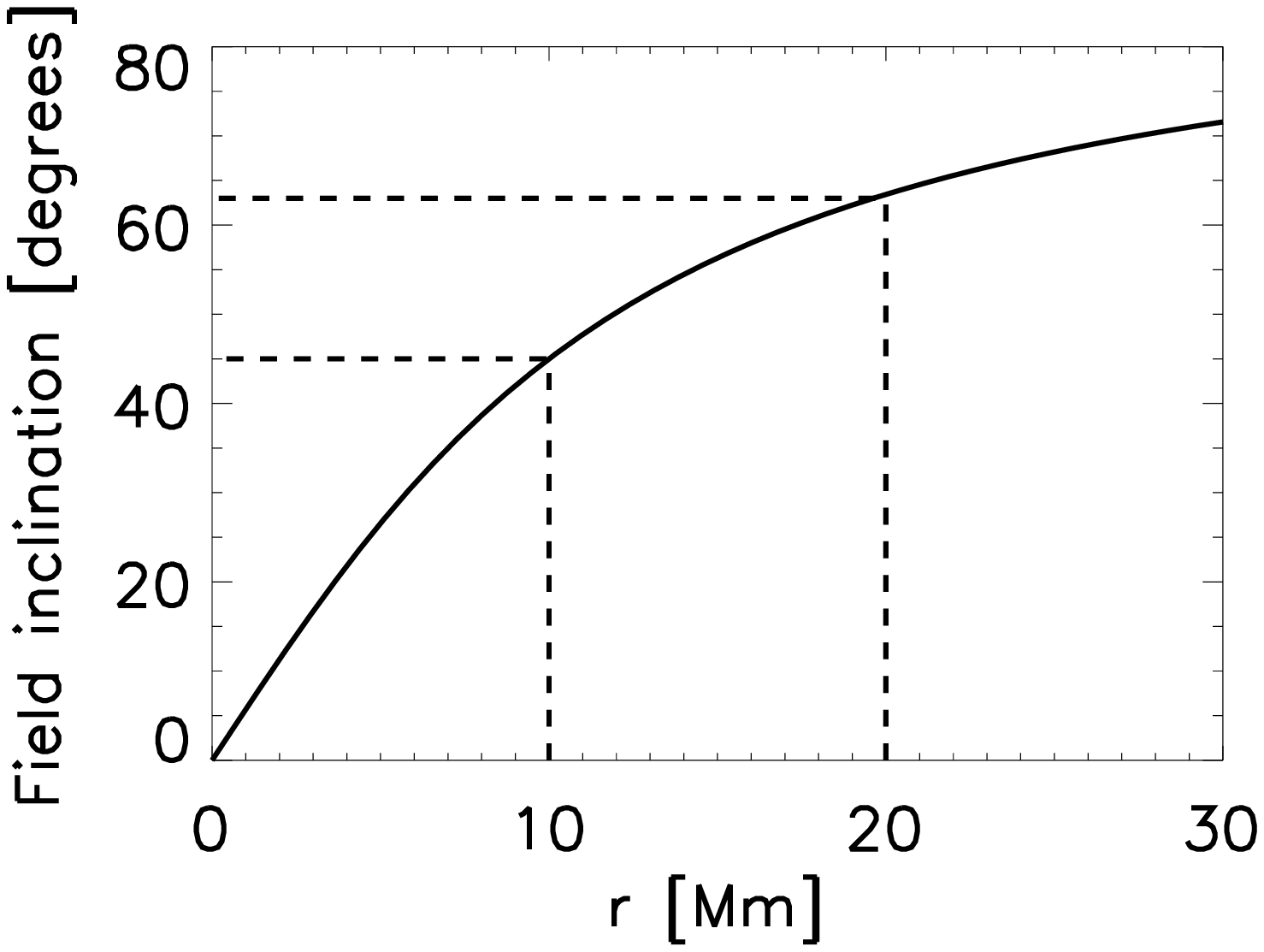}
\includegraphics[width=0.45\textwidth]{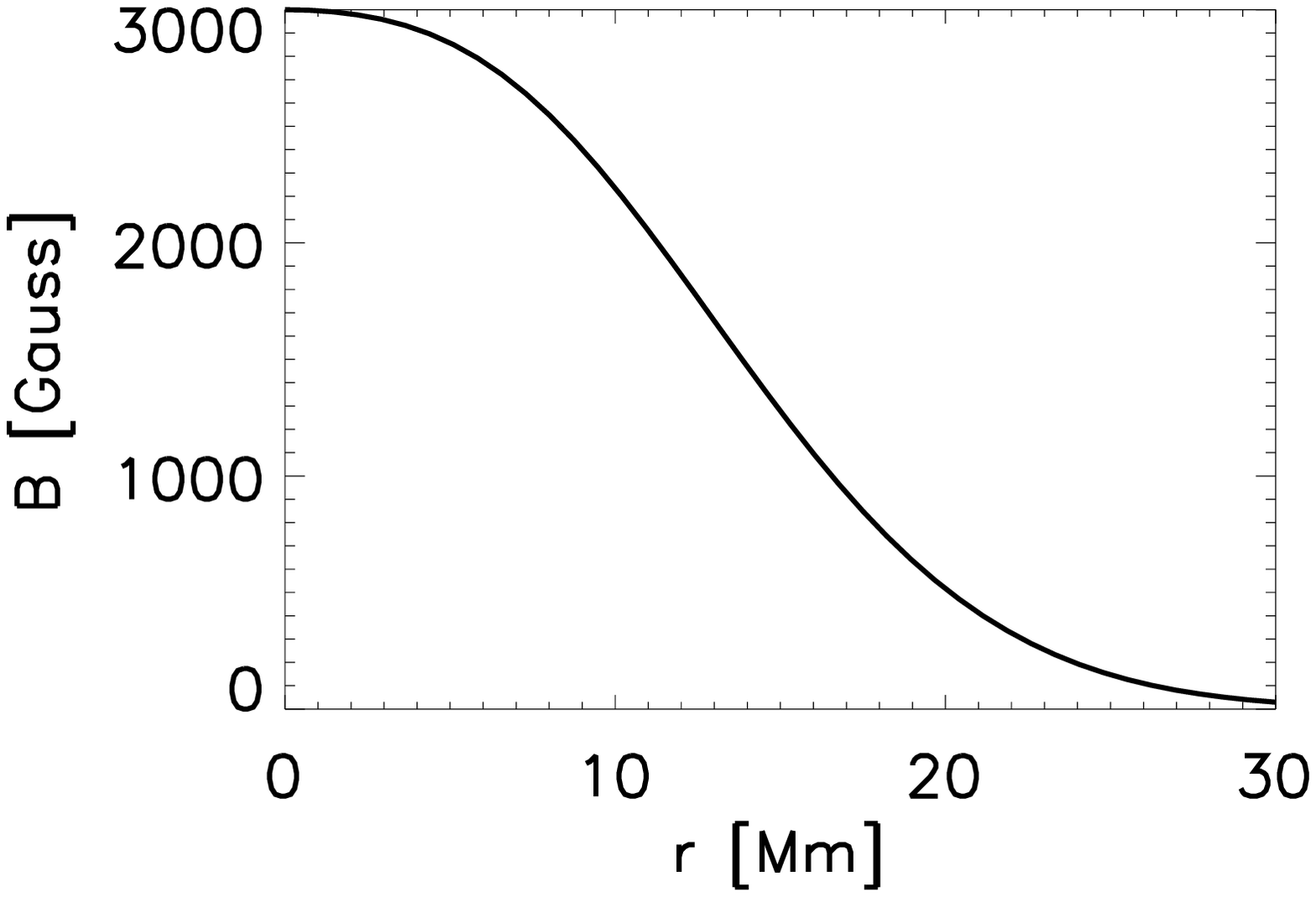}
\caption{The upper left panel shows the vertical magnetic field strength along the
axis of the spot. The upper right panel shows the ratio of the fast mode speed to the sound speed also along the
sunspot axis. The lower panels show the inclination (left) and strength (right) of the magnetic field at $z=0$  }
\label{fig_spot_field}
\end{figure}

Figure~\ref{fig_spot_field} summarizes the magnetic properties of the sunspot model.
The magnetic field strength increases rapidly with depth as a consequence of the choice of a large $\alpha_2$
in Equation~(\ref{eq.logistic}). Equation~(\ref{eq.radspot}) implies in turn that the radius of the tube shrinks fast,
from 10~Mm at the surface down to $1.3$~Mm at $z=-25$~Mm (just spatially resolved in the numerical simulations of Section~3).
This sunspot model---as monolithic as imaginable---is only one possible choice among many.   We note that it would be
straightforward to consider model sunspots that fan out as a function of depth, e.g., by decreasing the parameter $\alpha_2$.
Figure~\ref{fig_spot_field} also shows the ratio of the fast-mode speed, $v_{\rm f}$, to the sound speed, $c$, along the
sunspot axis, which is nearly unity for depths greater than 1 Mm, but increases very rapidly near the surface. For example,
we have $v_{\rm f}/c = 1.9$ at $z=z_u$ and $v_{\rm f}/c = 3.9$ at $z=-400$~km. The height at which the Alfv\'en and sound
speeds are equal ($a=c$) is $z=-750$~km.

We comment that the construction of both the thermodynamic and magnetic properties of this
model aims to get the surface properties correct. Additional information about the surface,
when available, could easily be incorporated into our sunspot model. Where information is missing
the assumption that the sunspot has similar properties to those of other sunspots is possibly the
best that can be done. A summary of the choices made to model the sunspot in AR 9787, which was
observed by MDI, is given in Table 1.

\begin{table*}
\center
\caption{Basic choices made in constructing the sunspot model for AR 9787.}
\begin{tabular}{ll}\hline
Umbral model &    \citet{maltby86} \\
Penumbral model& \citet{ding89}  \\
Umbral radius &  $r_{\rm{u}} = 10$ Mm  \\
Penumbral radius& $r_{\rm{p}} = 20$ Mm  \\
Umbral depression ($\tau_{5000}=1$)&  $z_0 - z_{\rm{u}}$ = 550 km    \\
Penumbral depression ($\tau_{5000}=1$)& $z_0 - z_{\rm{p}}$ = 150 km     \\
On-axis surface vertical field&   $B_{0}$ = 3 kG   \\
Radial profile of $B_z$ & Gaussian with HWHM=10 Mm \\
Field inclination at umbra/penumbra boundary & $45^\circ$  \\ \hline
\end{tabular}
\end{table*}

\section{Numerical simulation of the propagation of waves through the sunspot model}
We want to simulate the propagation of planar wave packets through the model sunspot
and surrounding quite-Sun using the SLiM code described in \cite{cameron07}
and \cite{cameron08}. The size of the box is $145.77$~Mm in both horizontal coordinates.
We used 200 Fourier modes in each of the two horizontal directions; almost all of the wave energy resides
in the 35 longest horizontal wavelengths.
The vertical extent of the box is $-25$~Mm $< z < 2.5$~Mm, where the physical domain, $-20$~Mm $< z < 0.5$~Mm, is sandwiched between two "sponge" layers that reduce strongly wave reflection from the boundaries \citep{cameron08}. We use 1098 grid points  and a finite difference scheme in the vertical direction.

Thus far the quiet-Sun reference model has been Model S which is convectively unstable.
For the purpose of computing the propagation of solar waves through our sunspot models using linear numerical simulations, we need a stable background model in which to embed the sunspot model. We use Convectively Stabilized Model B (CSM\_B) from \cite{schunker10}.
We keep the relative perturbations in pressure, density, and sound speed fixed. For example, this means that the new pressure is given by
\begin{eqnarray}
p(r,z)=P(r,z)   \;  \frac{P_{\mathrm{CSM\_B}}(z)}{P_{\mathrm{qs}}(z)},
\end{eqnarray}
where $P(r,z)$ is the pressure as discussed above. The same treatment is applied to density, $\rho$, and sound speed, $c$.

Separate simulations are done for f, p$_1$, and p$_2$ wave packets. In each case, a wave
packet is made of approximately 30 Fourier modes. At $t=0$ all the modes are in phase at $x=-43.73$ Mm,
while the sunspot is centered at $x=y=0$. The initial conditions at $t=0$ are such that the wave packet
propagates in the positive $x$ direction, towards the sunspot.

The Alfv\'en velocity increases strongly above the photospheric layers of the spot
(see Figure~\ref{fig_spot_field}). This is a problem for two reasons. The first and most important
of these is that it causes the wavelengths of both the fast-mode and Alfv\'en waves to become
very large. The upper boundary, which is located at $z=2.5$ Mm, becomes only a fraction of a
wavelength away from the surface. This makes the simulation sensitive to the artificial upper
boundary. The second reason is that our code is explicit and hence subject to a CFL condition:
High wave speeds require very small time steps and correspondingly large amounts of
computer power.  In order to address this problem, we notice that we expect all solar waves
that reach the region where the Alfv\'en speed is high to continue to propagate upward out of the
domain. We therefore reduce the Alfv\'en speed in this region in the simulation to increase the
time the waves spend in the sponge layer. This reduces the influence of the upper boundary and allows us to have
a larger time step.   In practice, we multiply the Lorentz force by $8 c^2 /(8 c^2 +a^2)$, where
$a(r,z)$ is the Alfv\'en speed, and $c(z)$ is the sound speed of the quiet Sun at the same height.
This limits the fast-mode speed to be a maximum of three times the quiet-Sun
sound speed at the same geometric height. Since the sound speed near the
$\tau=1$ layer in the spot is less than half of that of the quiet Sun at the same height,
the maximum value of the fast mode speed in this critical region is approximately six times the local
sound speed. The functional form chosen to limit the Lorentz force modifies $a$ at and below
the $c=a$ level by less than 2\%. This corresponds to a change in field strength of
less than 30~G or a minimal change in the height of the  $c=a$ surface.

\section{Preliminary comparison of simulations with observed MDI cross-covariances of the
Doppler velocity}

There exist excellent observations (SOHO/MDI Doppler velocity) of helioseismic waves around the sunspot in AR 9787
\citep{gizon09}. The wave field around the sunspot can be characterized by the temporal cross-covariance
of the observed random wave field. It has been argued \citep[][and references therein]{Gizon10} that the observed cross-covariance
is closely related to the Green's function (the response of the Sun to a localized source).
Hence the observed cross-covariance is comparable to the surface
vertical velocity from initial-value numerical simulations of the wave packet propagation.
In \cite{cameron08} we studied the propagation of an f-mode wave packet through a
simplified magnetohydrostatic sunspot model. We found that we could constrain the
surface magnetic field by comparing the observed f-mode cross-covariance with a simulated f-mode wave packet.

Here we assess the seismic signature of the semi-empirical sunspot model described in Section~2 by comparing the simulations and observations.
The cross-covariance is constructed in the same way as in \cite{cameron08}, to which we refer the reader.  In short, it is computed according to
\begin{equation}
C(x,y,t)=\int_0^T {\bar{\phi}} (t')\phi(x,y,t'+t) \, dt',
\end{equation}
where $T=7$~days is the observation time, $\phi$ is the observed Doppler velocity, ${\bar{\phi}}$ is the average of $\phi$ over the line $x=-43.73$~Mm, and $t$ is the time lag. We select three different wave packets (f, p$_1$, and p$_2$) by filtering $\phi$ along particular mode ridges.
As said in Section~3,  for the numerical simulations we consider plane wave packets starting at $x=-43.73$~Mm and propagating in the $+x$ direction towards the model sunspot (at the origin). The initial conditions of the simulation are chosen such that, in the far field, the simulated vertical velocity has the same temporal power as the observed cross-covariance.

\begin{figure}
\includegraphics[width=0.45\textwidth, trim=2.5cm 2.5cm 3.5cm 0.05, clip=true]{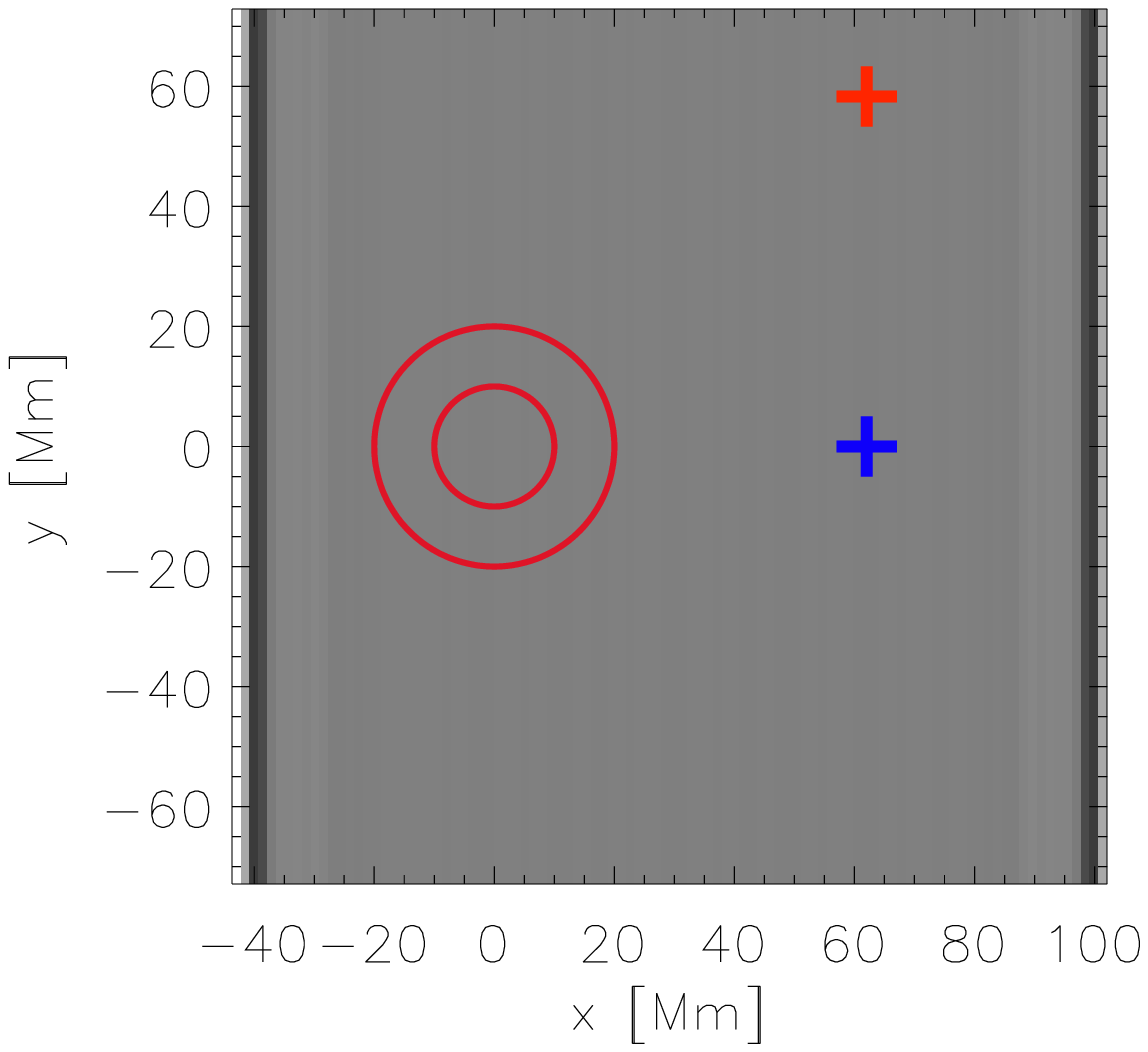}
\includegraphics[width=0.45\textwidth, trim=2.5cm 2.5cm 3.5cm 0.05, clip=true]{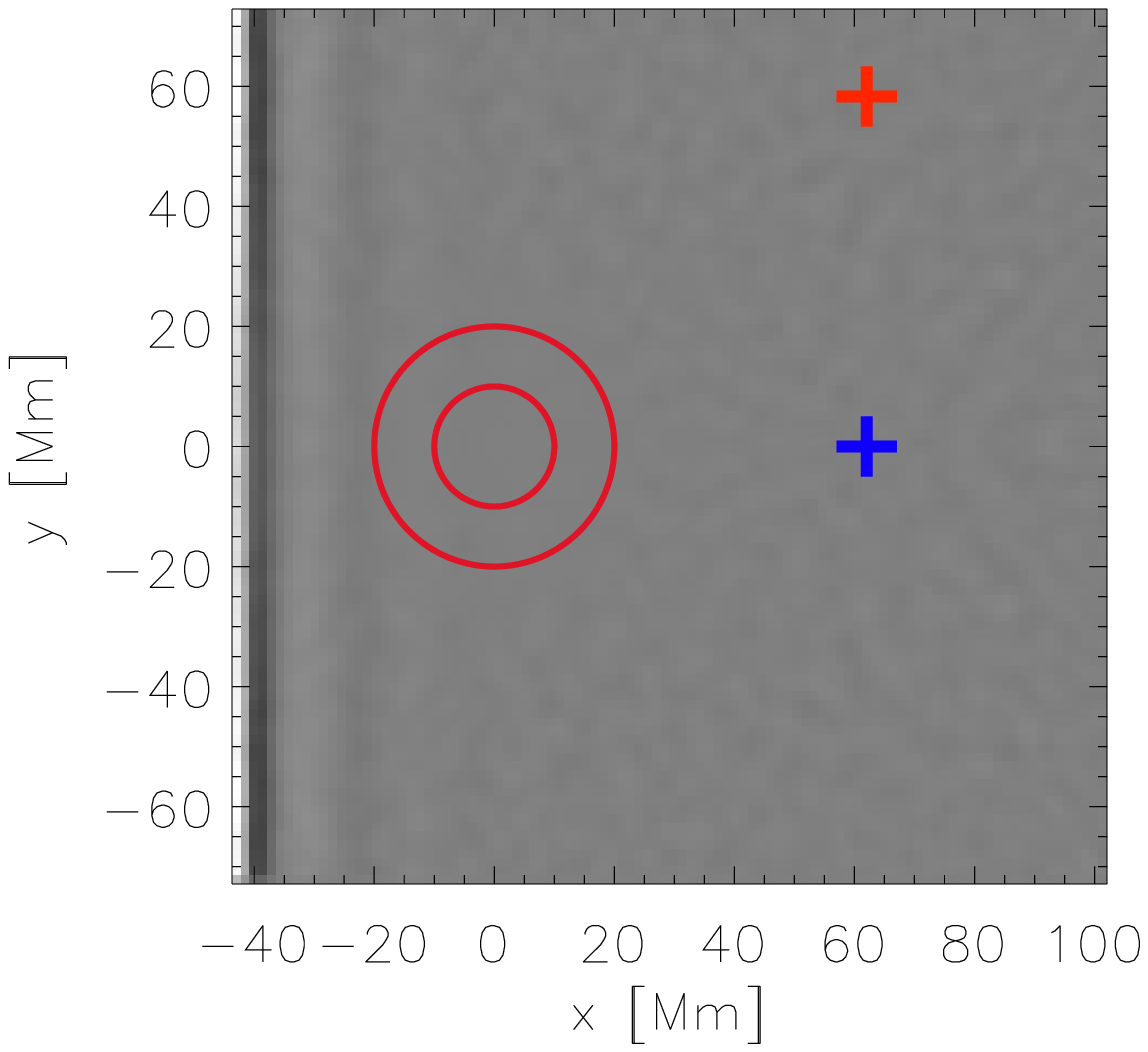}

\includegraphics[width=0.45\textwidth, trim=2.5cm 2.5cm 3.5cm 0.05, clip=true]{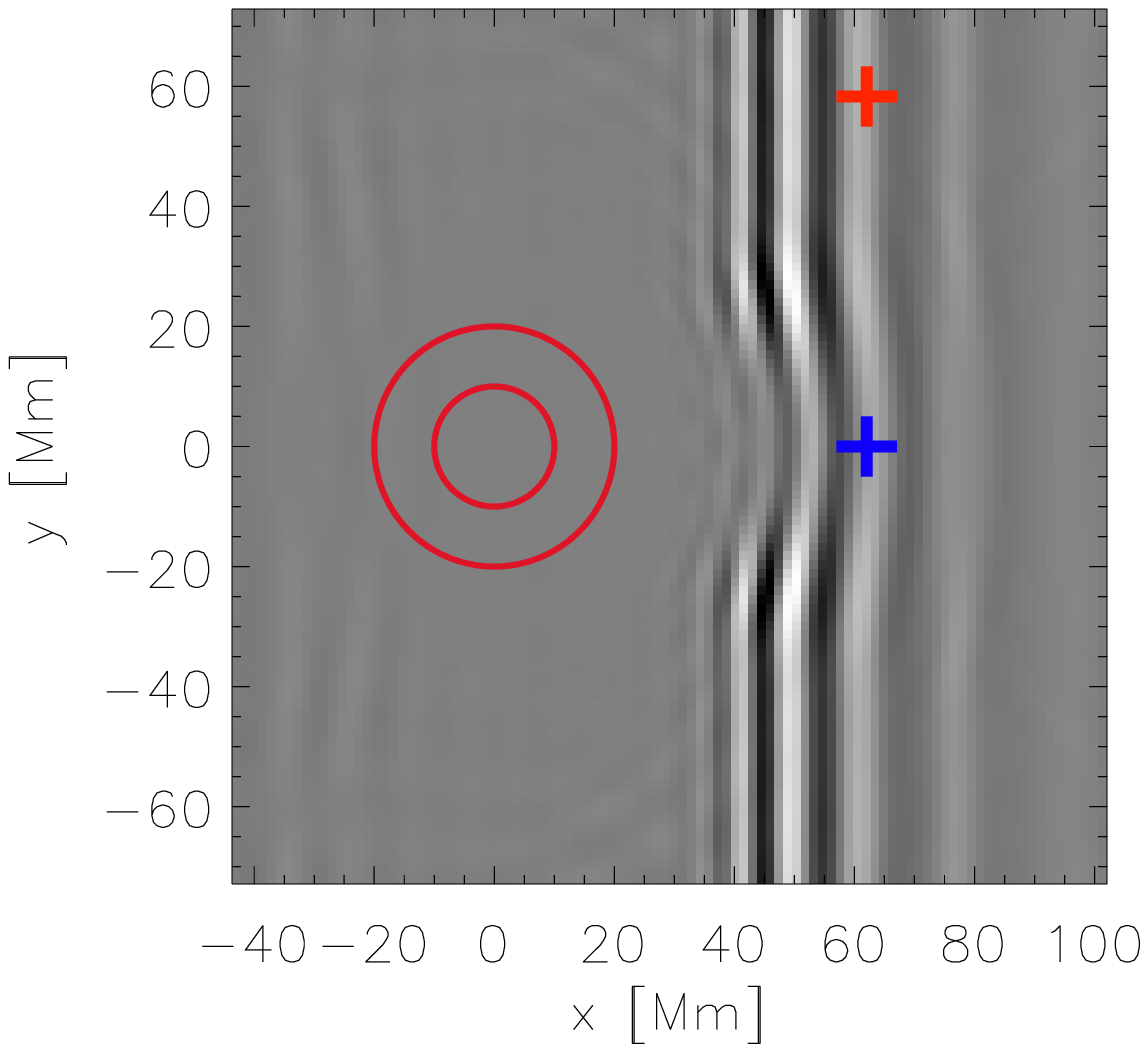}
\includegraphics[width=0.45\textwidth, trim=2.5cm 2.5cm 3.5cm 0.05, clip=true]{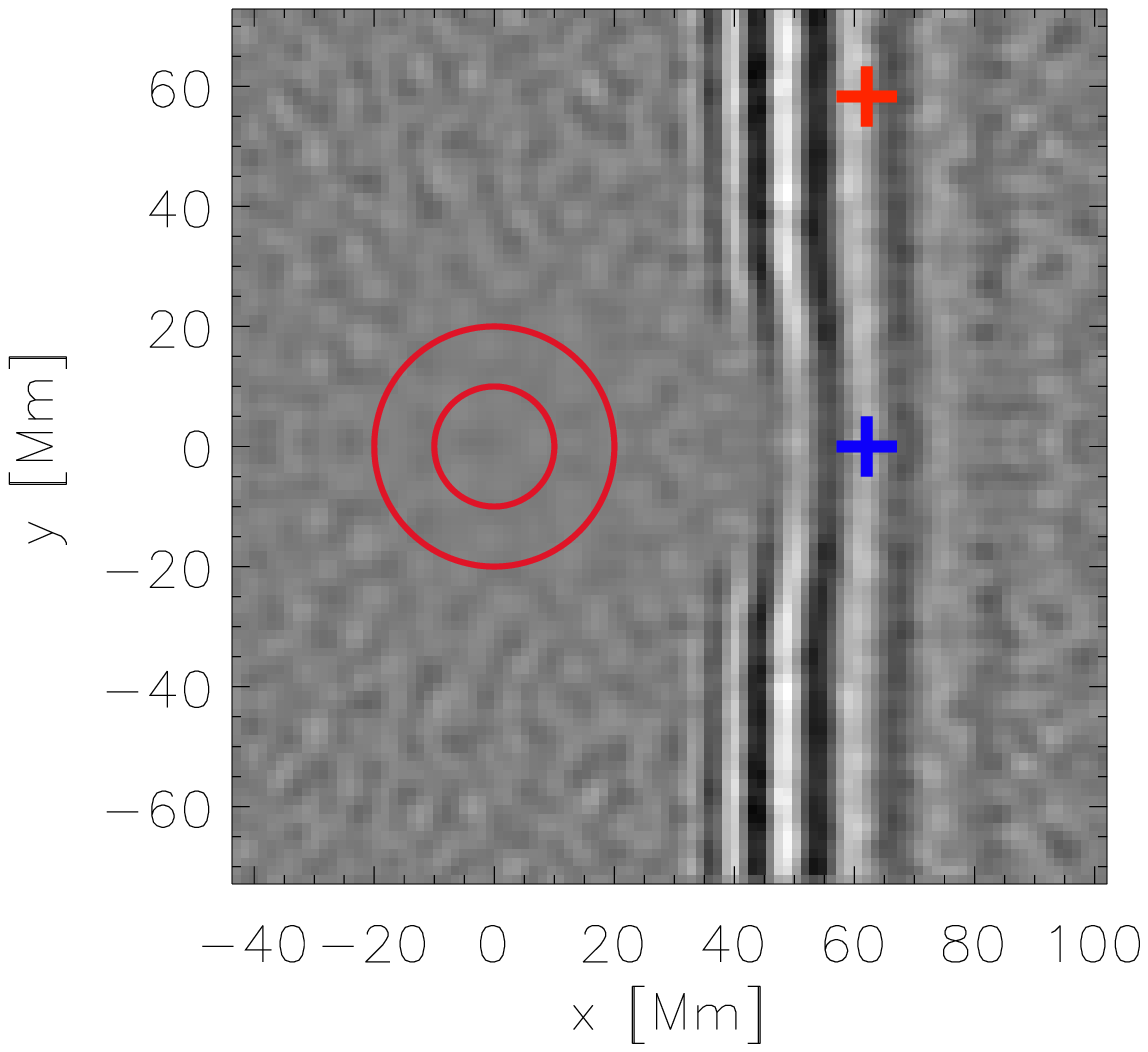}

\includegraphics[width=0.45\textwidth, trim=2.5cm 2.5cm 3.5cm 0.05, clip=true]{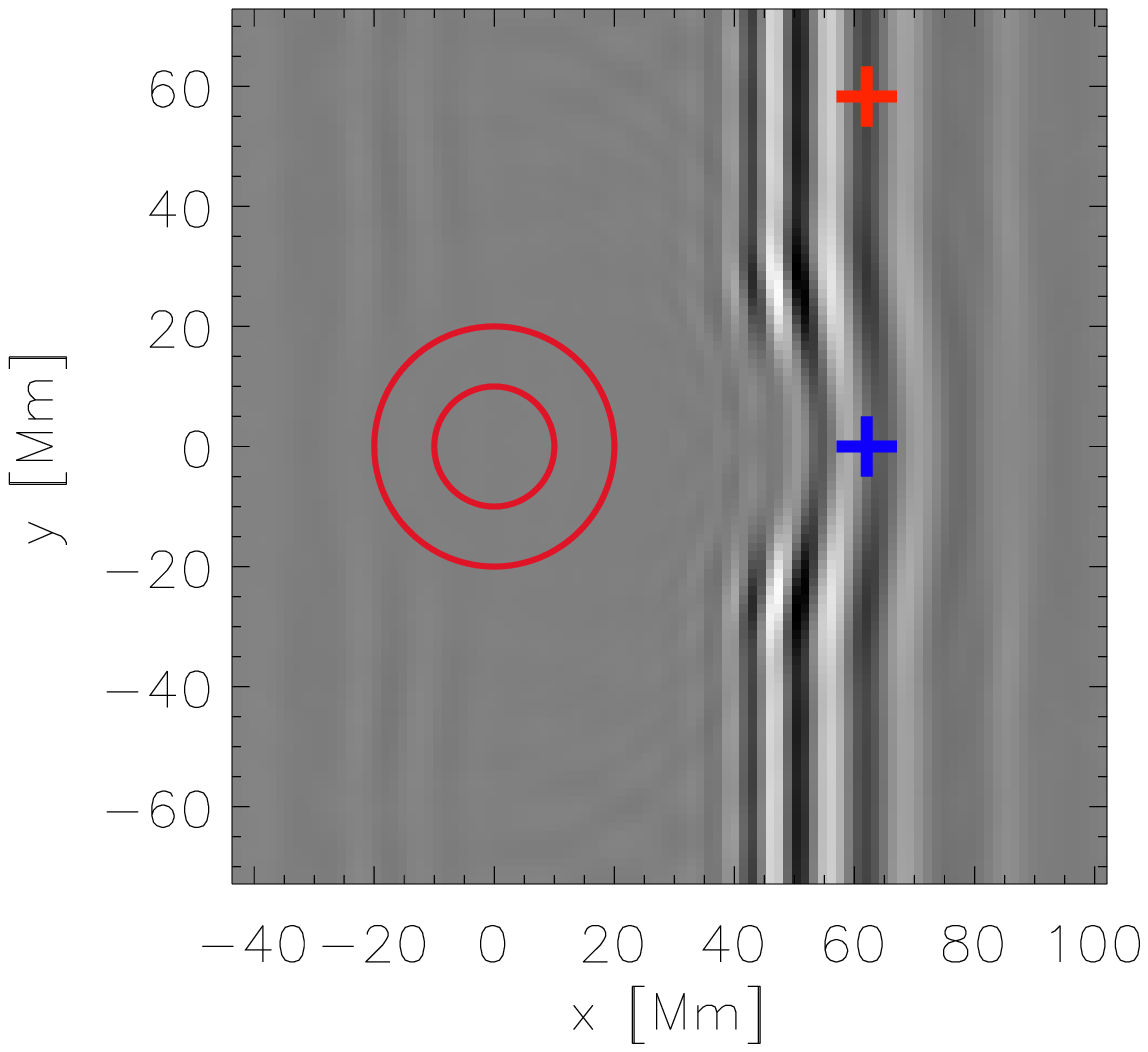}
\includegraphics[width=0.45\textwidth, trim=2.5cm 2.5cm 3.5cm 0.05, clip=true]{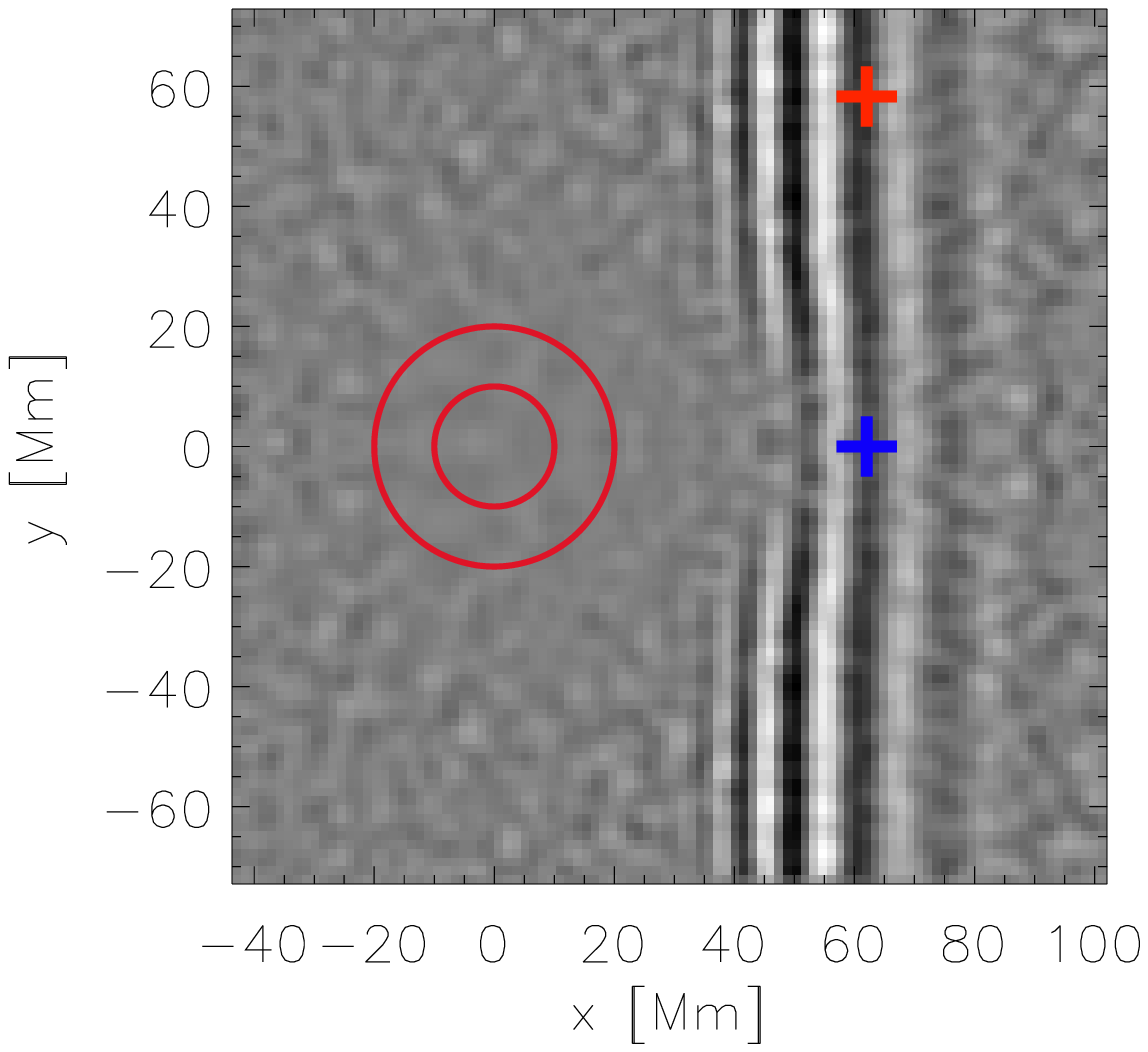}

\includegraphics[width=0.45\textwidth, trim=2.5cm 0.05 3.5cm 0.05, clip=true]{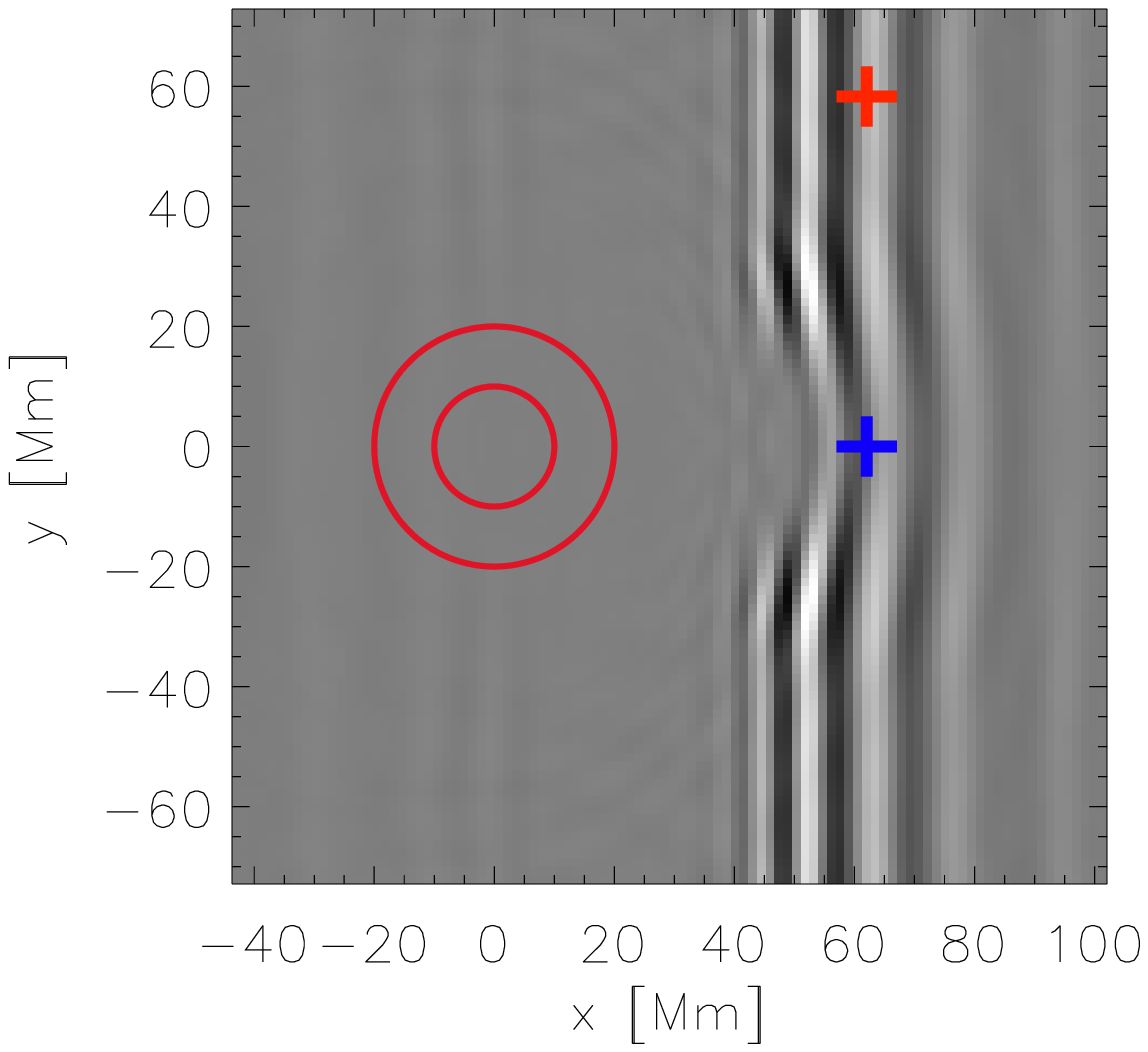}
\includegraphics[width=0.45\textwidth, trim=2.5cm 0.05 3.5cm 0.05, clip=true]{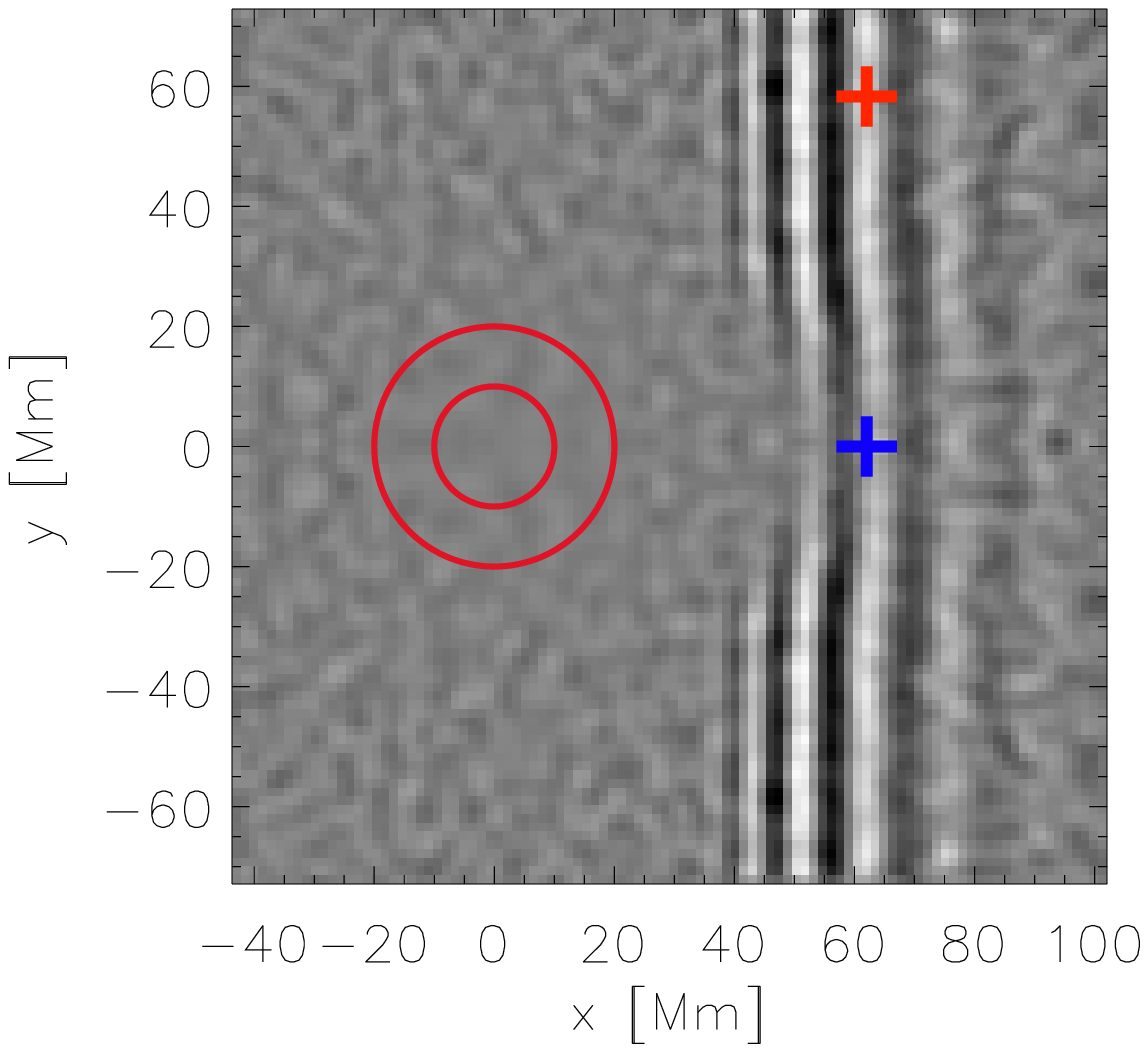}

\caption{(Left panels) Simulated vertical velocity of p$_1$ wave packets at times $t=0,$ 140, 144, and
148~min (from top to bottom). White shades are for positive values, dark shades for negative values.
The gray scale covers the full range of values in each row.
(Right panels) Observed cross-covariance at the corresponding time-lags.
The $t=0$ frame from the simulation is the prescribed initial condition.
The red and blue crosses (hereafter points A and B respectively) indicate two particular locations used in the subsequent analysis (Figures~\ref{fig_ts} and \ref{fig_fd}).}
\label{fig_p1_s}
\end{figure}

Figure~\ref{fig_p1_s} shows the observed cross-covariances
and simulated wave packets for the p$_1$ modes at four consecutive time lags.
The main features seen in the cross-covariances, the speedup of the waves across the
sunspot as well as their loss of energy at short wavelengths, are also seen in the simulations.
The details of the perturbed waveforms can only be understood in the context of finite-wavelength scattering \citep[][]{Gizon2006}.

\begin{figure}
\includegraphics[width=0.45\textwidth]{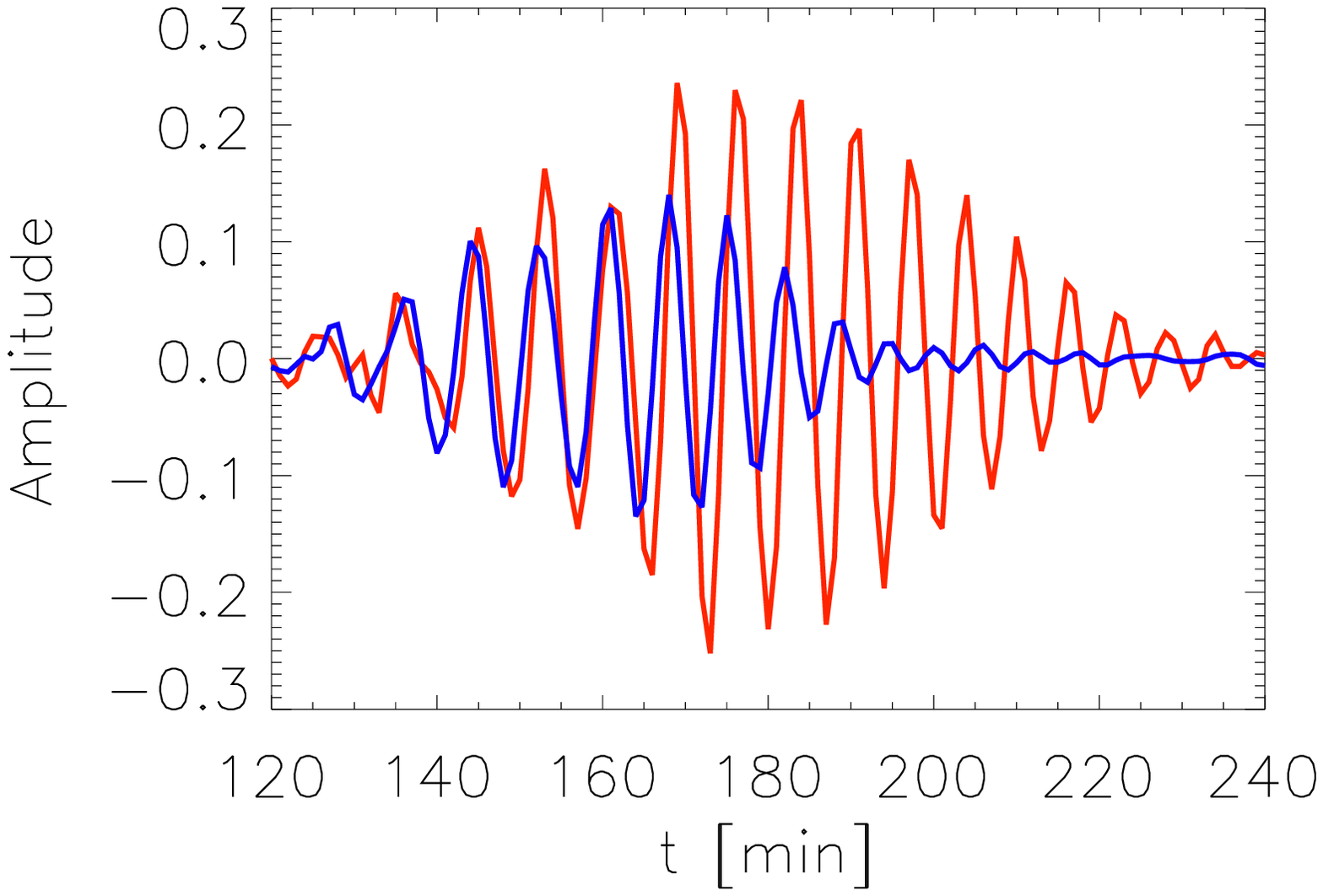}
\includegraphics[width=0.45\textwidth]{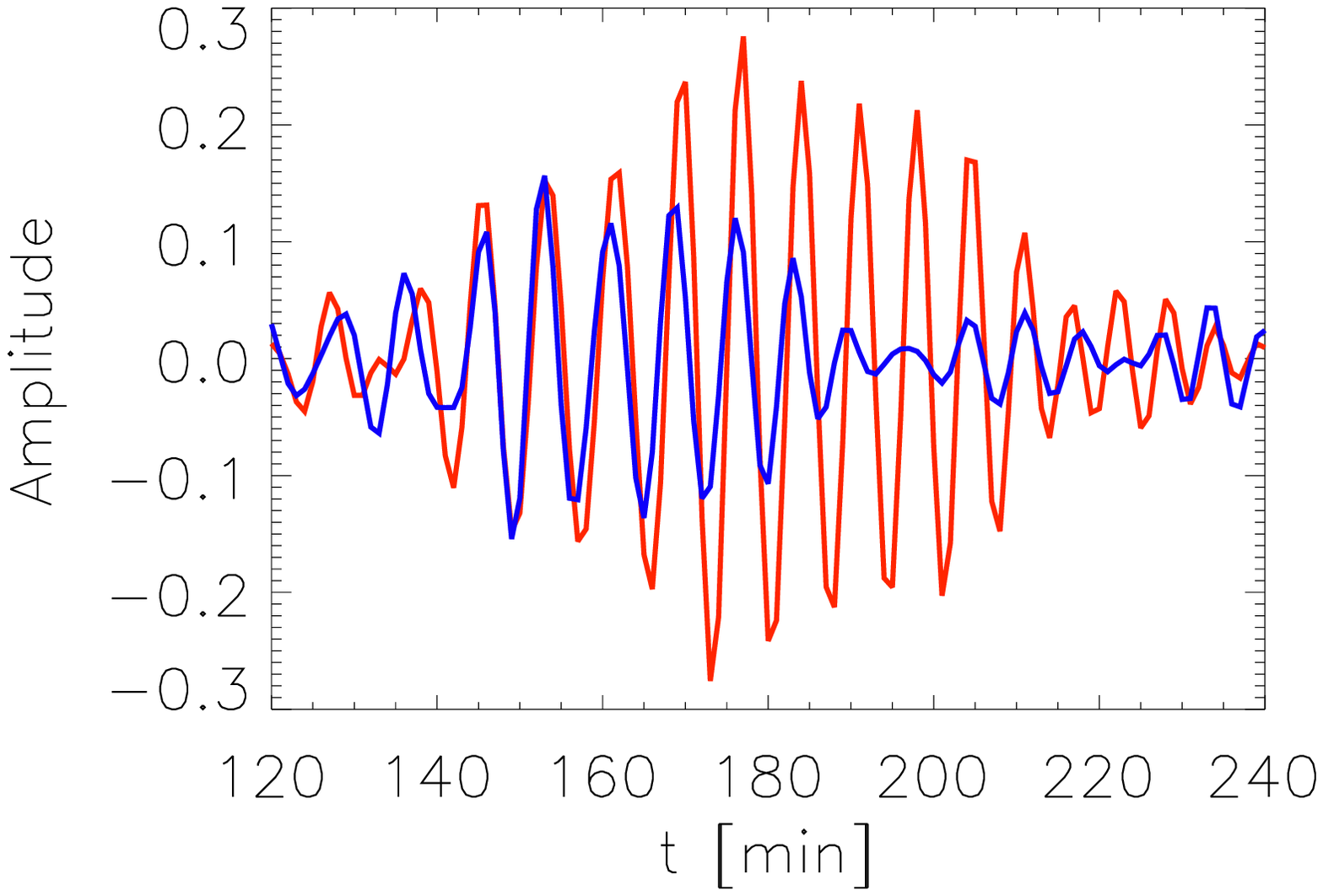}

\includegraphics[width=0.45\textwidth]{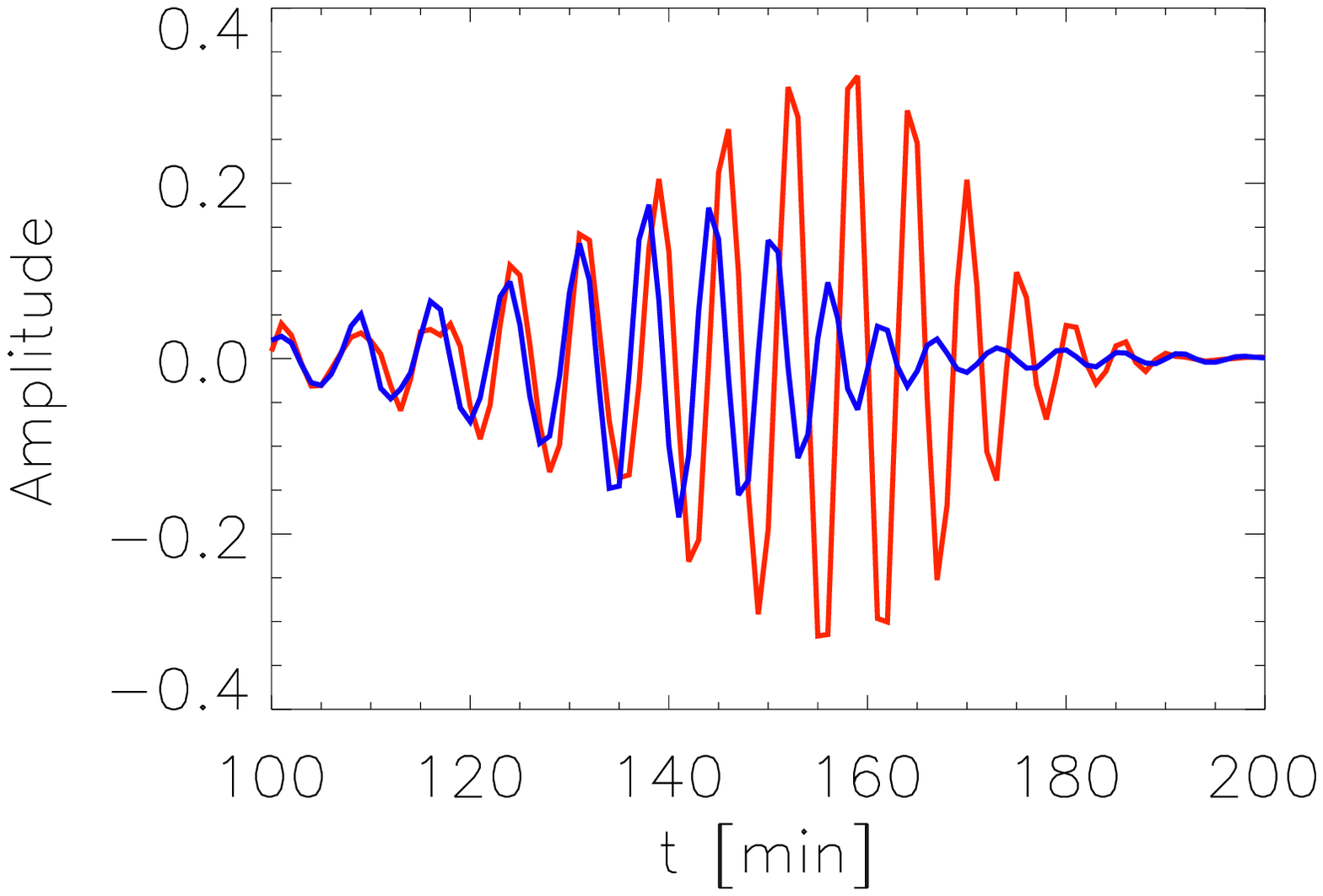}
\includegraphics[width=0.45\textwidth]{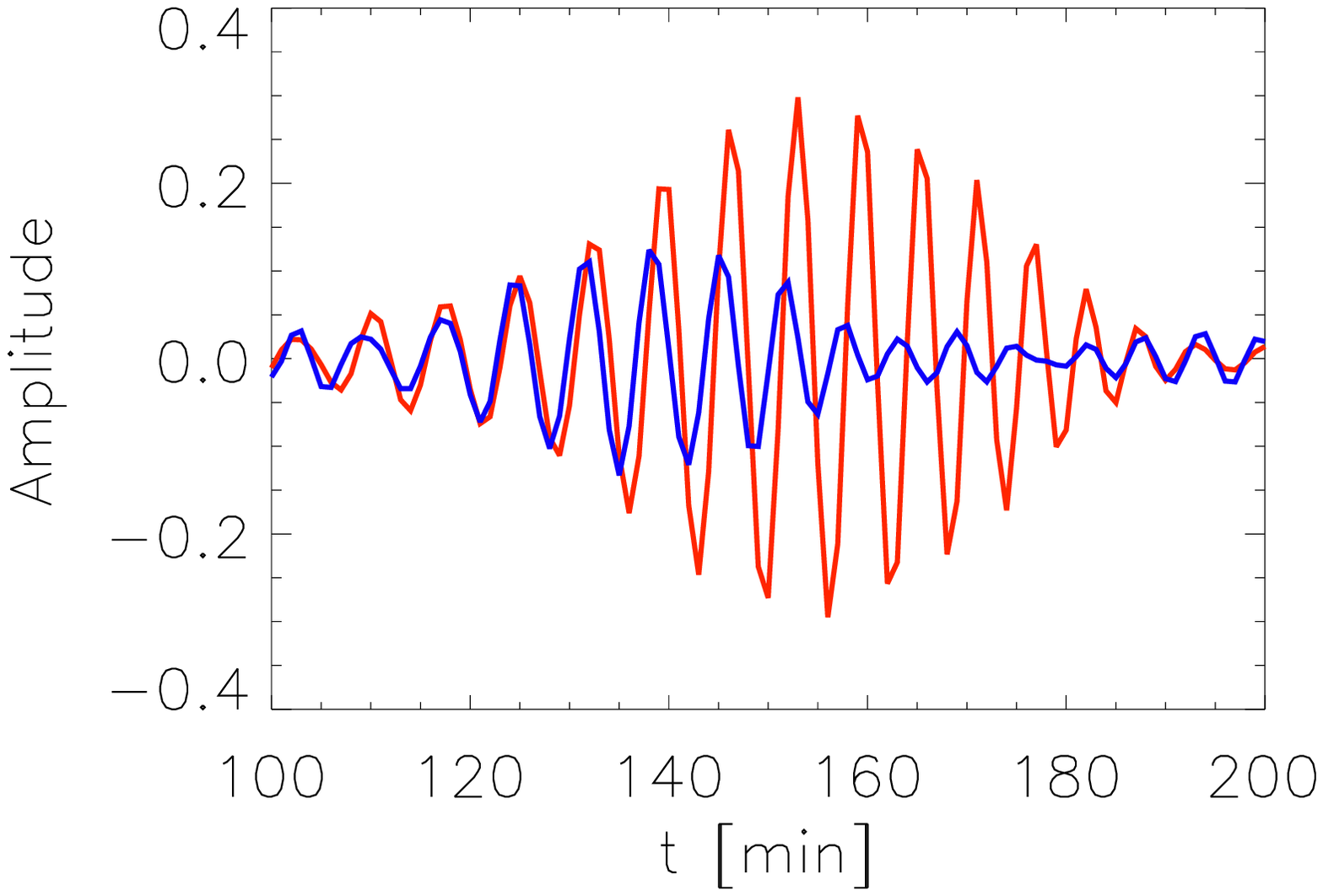}

\includegraphics[width=0.45\textwidth]{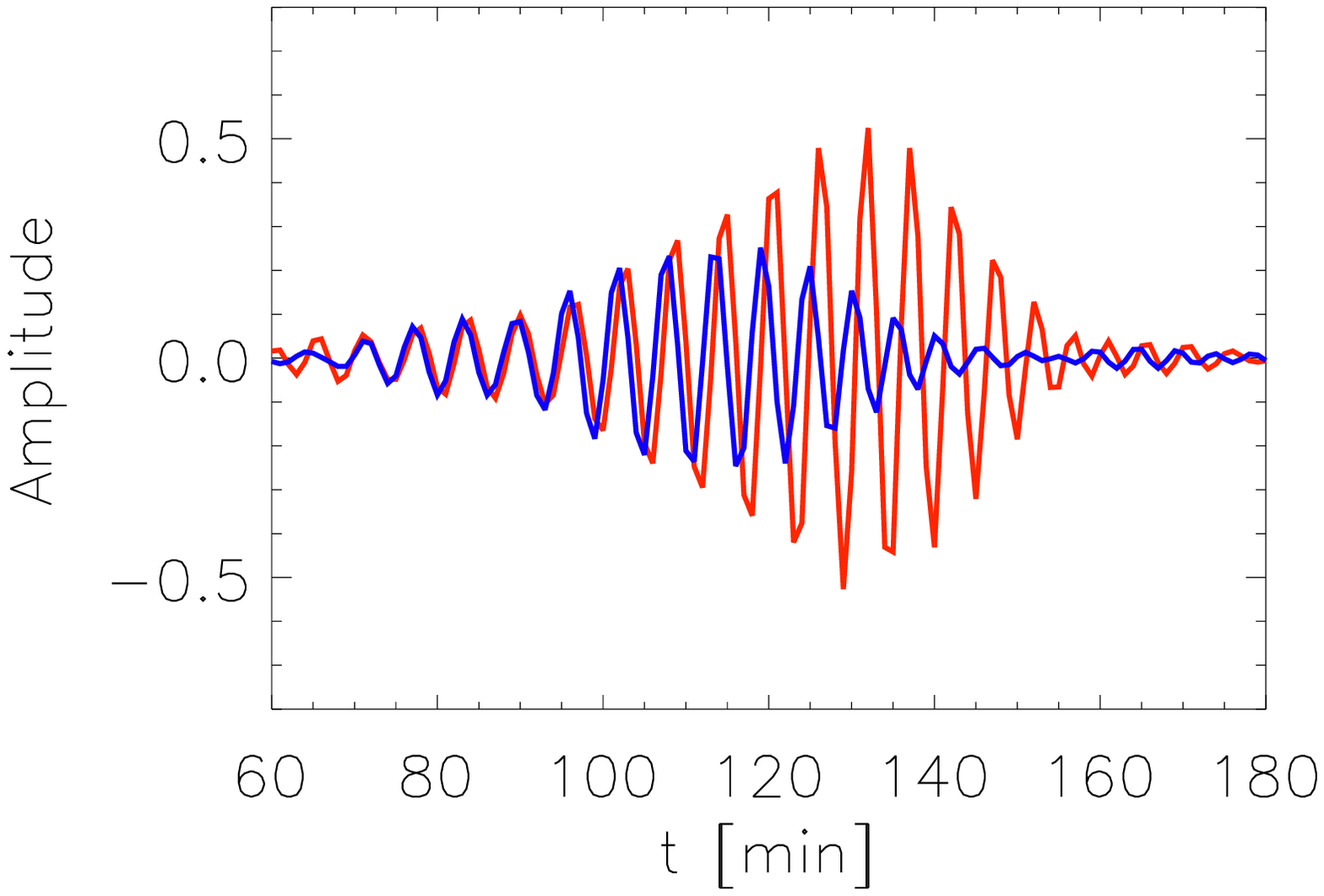}
\includegraphics[width=0.45\textwidth]{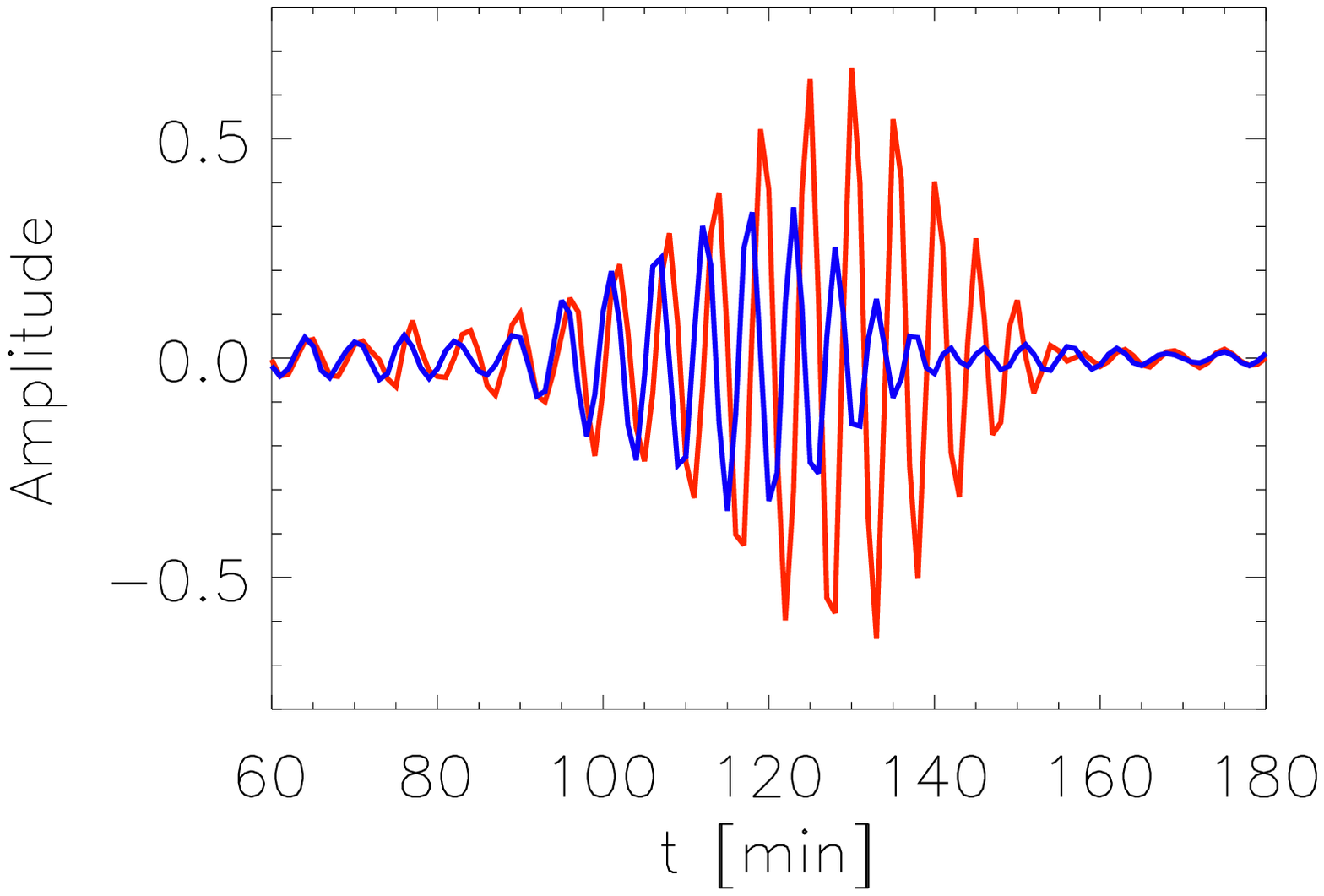}
\caption{Time series from the simulations (left panels) or from the
observed cross-covariance
(right panels) for the (top to bottom) f, p$_1$ and p$_2$ wave packets. The red curve corresponds
to point A (the red cross) in Figure~\ref{fig_p1_s}, the blue curve to
point B (the blue cross).}
\label{fig_ts}
\end{figure}

For a more detailed comparison we have concentrated on the two spatial locations $A=(x,y)=(60,0)$~Mm and $B=(60,60)$~Mm, marked
with crosses in Figure~\ref{fig_p1_s}. The first point lies behind the sunspot, where the effects of the sunspot are easily noticeable.
The second point is away from the scattered field and serves as a quiet-Sun reference.
The corresponding simulated and observed wave packets are plotted as functions of time in Figure~\ref{fig_ts}. The match again looks qualitatively good.

\begin{figure}
\includegraphics[width=0.45\textwidth]{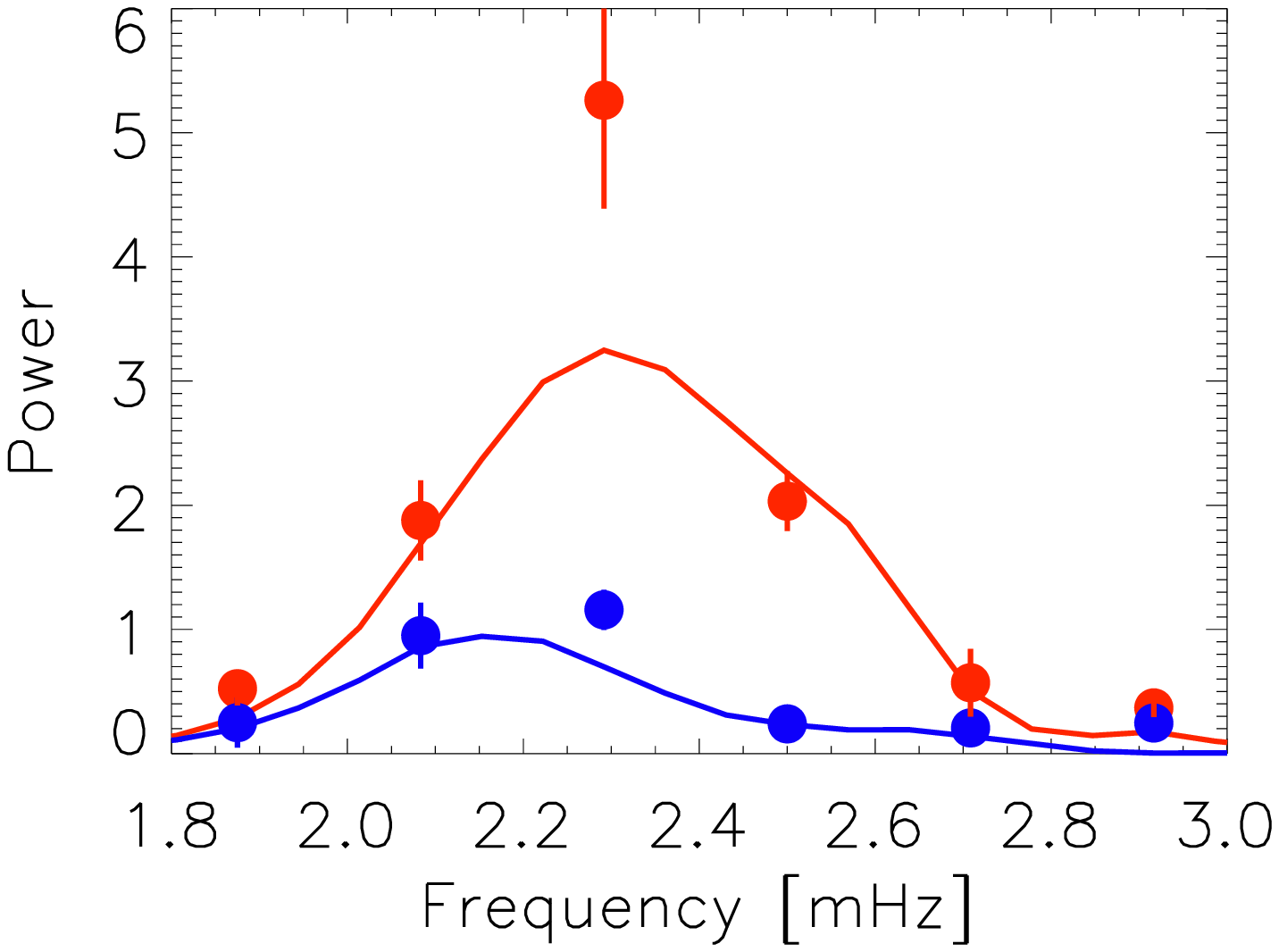}
\includegraphics[width=0.45\textwidth]{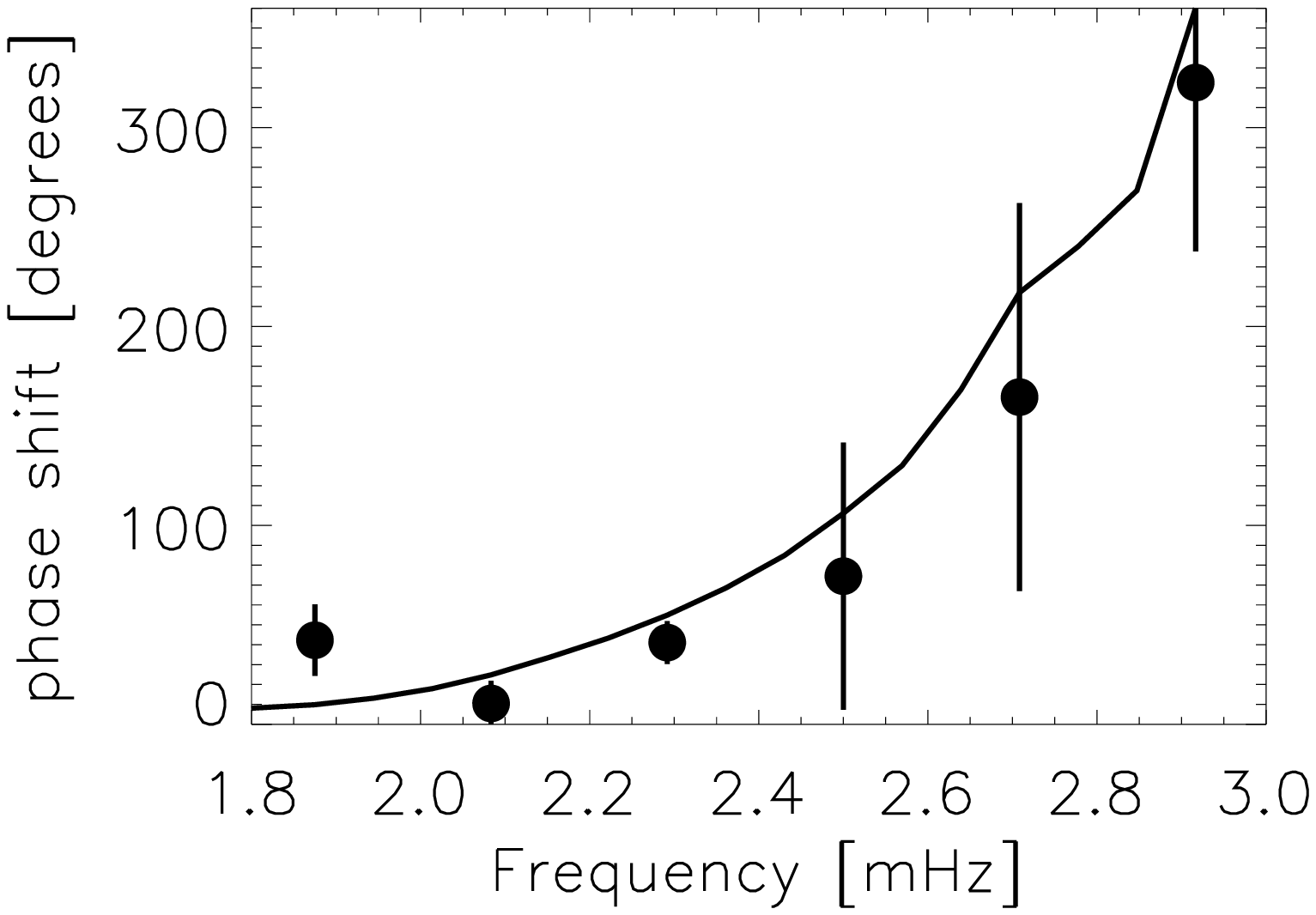}

\includegraphics[width=0.45\textwidth]{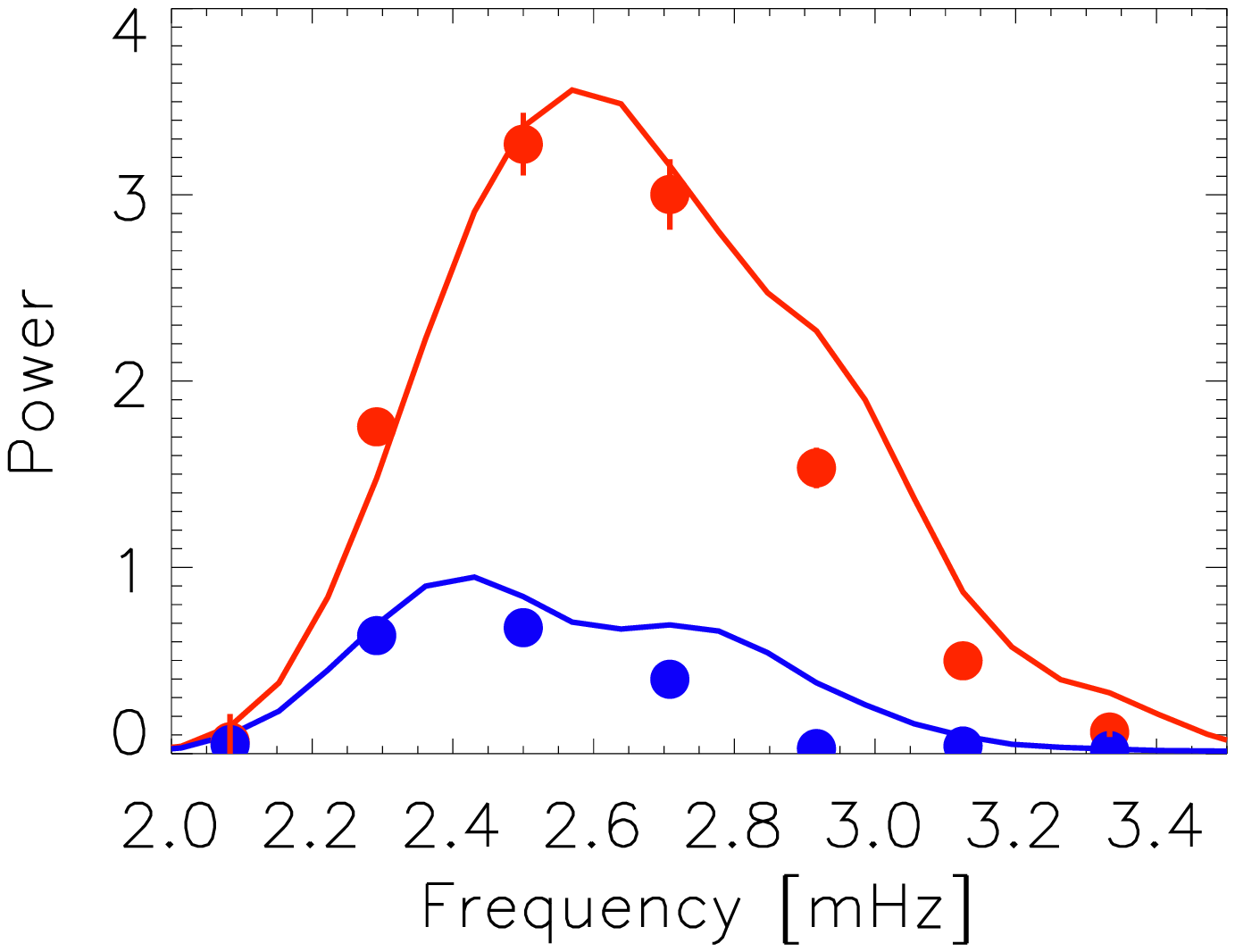}
\includegraphics[width=0.45\textwidth]{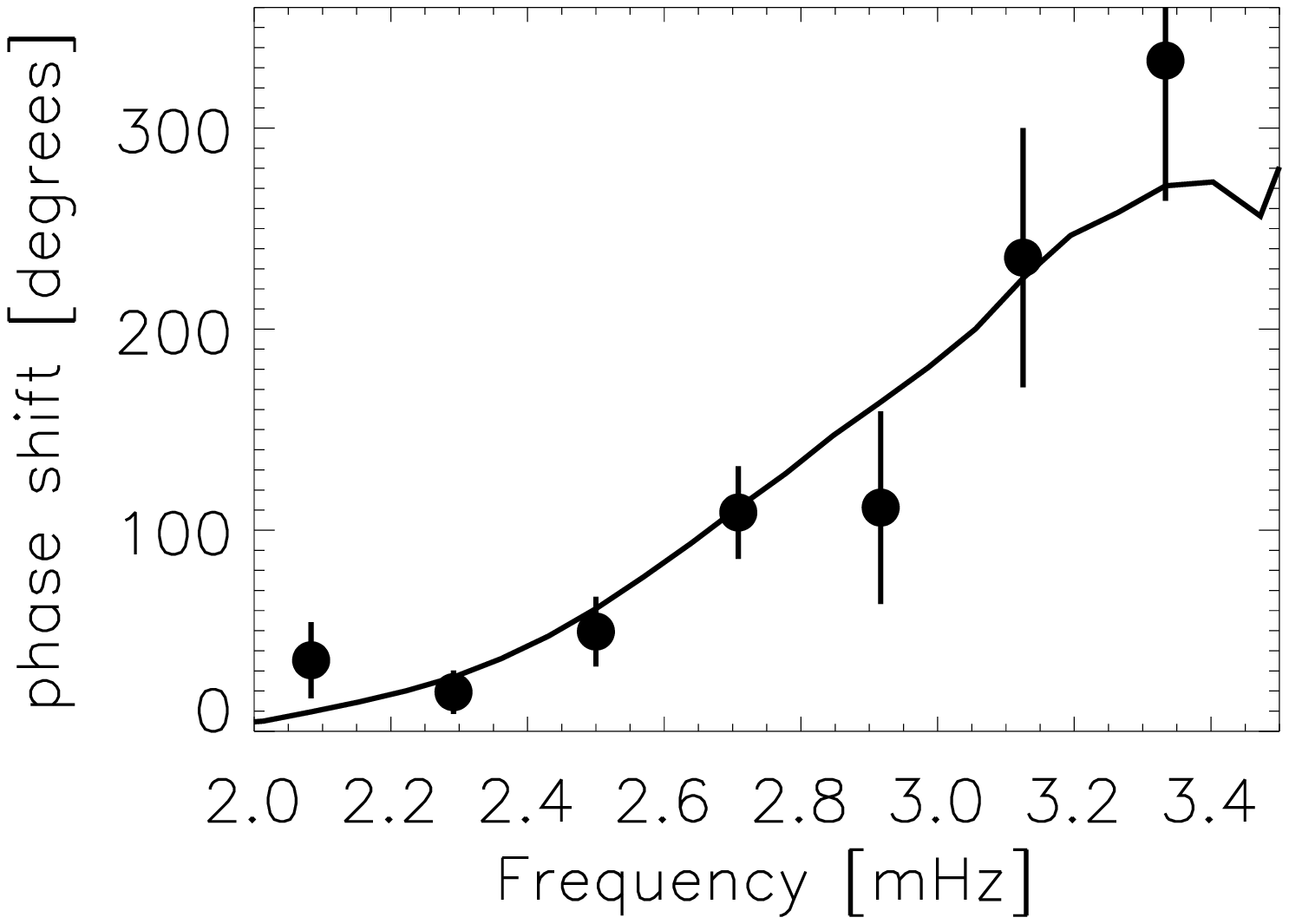}

\includegraphics[width=0.45\textwidth]{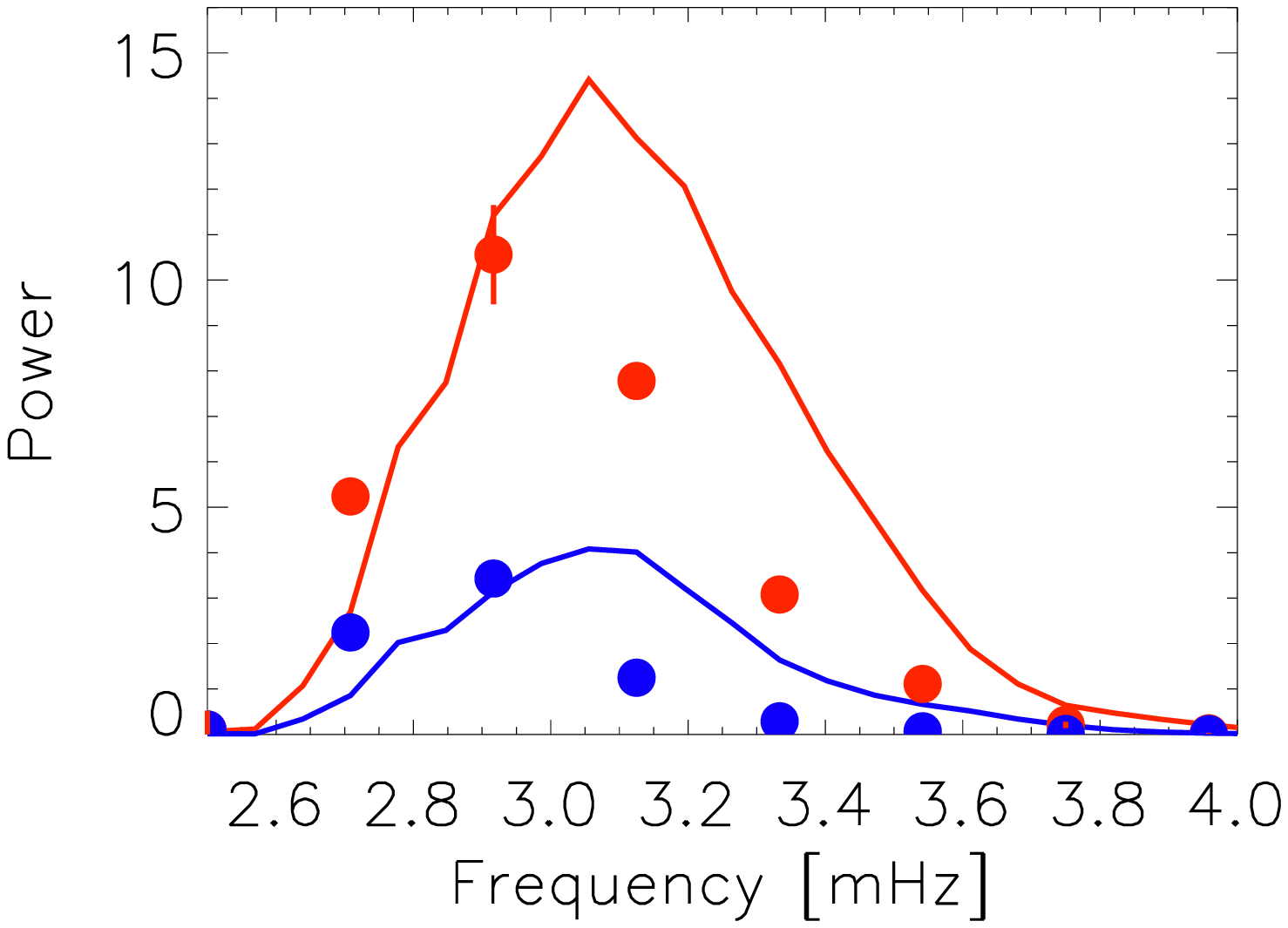}
\includegraphics[width=0.45\textwidth]{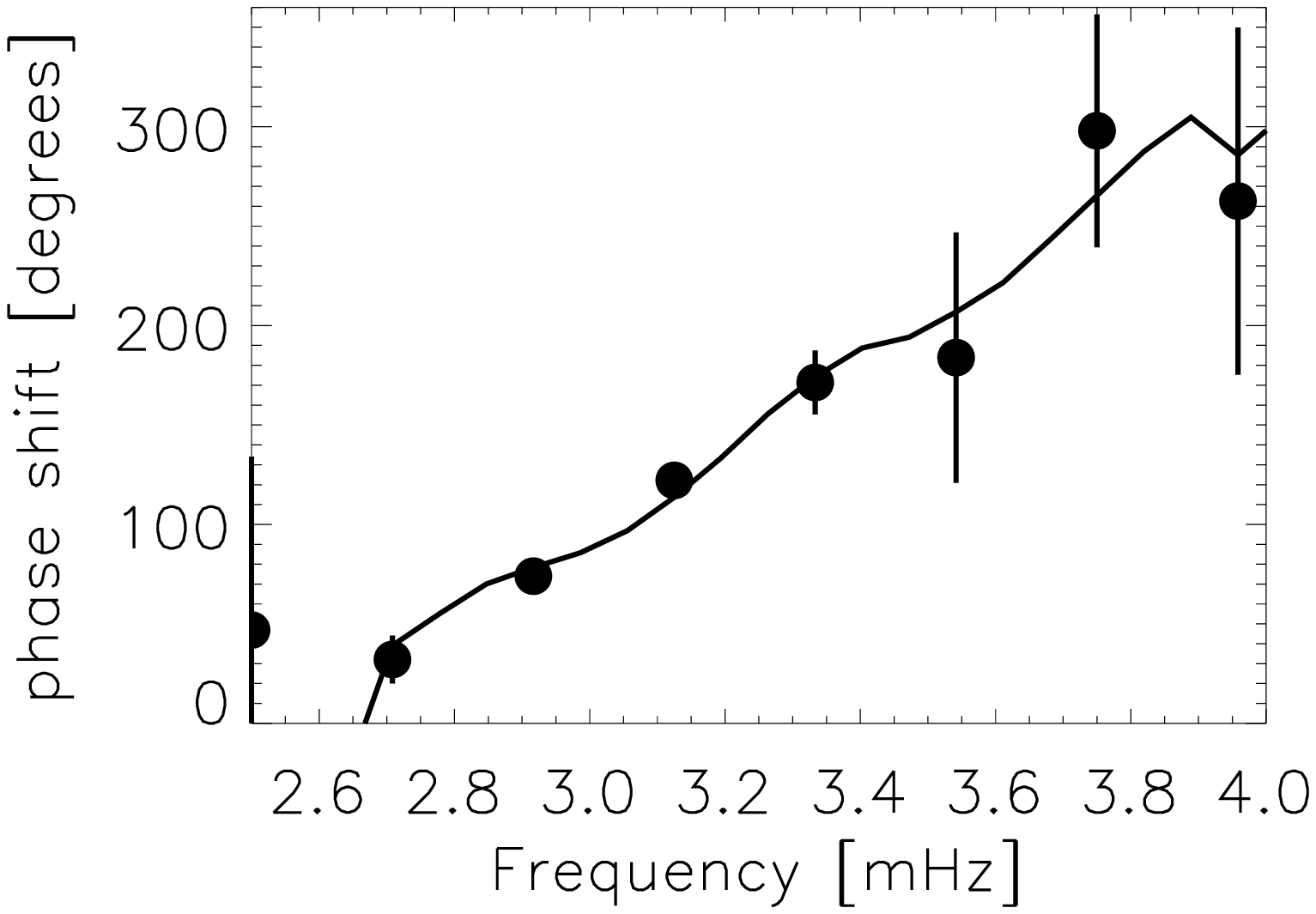}
\caption{Frequency analysis of the curves shown in Figure 8.
The left panels shows the power spectra of the time series from Figure 8,
with the solid lines indicating the values for the numerical simulations
and the dots indicating the values from the observed cross-covariances.
The red curves are for point B (away from the sunspot) abd the blue curves for point A (behind the sunspot).
The top panels are for f mode wave packets, the middle for p$_1$, and the bottom for p$_2$.
The right panels show the phase differences between points A and B, i.e. sunspot minus quiet-Sun.
The solid lines are for the numerical simulations and the dots are for the observations.
The error bars are root-mean-square deviations computed over $0.3$-mHz intervals.
}
\label{fig_fd}
\end{figure}

To proceed further we Fourier analyze the time series
at points A and B: we consider the power spectra and phases.
The temporal power spectra are shown in the left panels of Figure~\ref{fig_fd}.
The power spectrum of the simulations is similar to that of cross-covariance for
the p$_1$ modes, less so for the f and p$_2$ modes. We note that the differences between
the simulated and observed power spectra are seen at both spatial points.
We also analyzed the phase difference between the wave packets at the two points (A minus B) as a function of
frequency (Figure~\ref{fig_fd}). The phase shifts introduced by the sunspot are reasonably well reproduced by the
simulations, although some differences exist especially for the f modes.

In both the observations and the simulations we see that the power in the f and p$_1$ wavepackets
peaks substantially below $3$~mHz. The reason for this can be understood by noting that
wave attenuation increases sharply with frequency. Even though the initial power spectra have most of their energy near 3~mHz,
 very little of the power at and above 3~mHz reaches the points $A$ and $B$. The remaining low frequency
waves have more sensitivity to the deeper layers and are thus the most important in
determining the subsurface structure of the sunspot. It is also apparent that the sunspot more
strongly 'damps' higher frequency waves. This is understandable since these waves have more of their
energy in the near-surface region, where mode conversion (from the f and p modes into
Alfv\'en and slow-magnetoacoustic modes that propagate away from the surface) is efficient.
Over the frequency range studied, the power 'absorption' coefficient increases with frequency \citep[in accordance with][]{Braun95,Braun2008}.
The same trend is also seen in the phase shifts, with low frequency waves being almost unaffected by the sunspot whilst high
frequency waves have phase shifts which range up to 300 degrees (for the frequencies shown). A strong frequency dependence of the travel-time perturbations had already been reported by \citet{Braun2008} for ridges p$_1$ through p$_4$ \citep[see also][]{Braun2006,Birch2009}.
The reason why the phase shifts are strongest for high-frequency  waves is again that they have more energy near the surface.
The possibility of such strong phase shifts should be born in mind when interpreting (and measuring) helioseismic waves
near sunspots.

\section{Conclusion}
We have outlined a method for constructing magnetic models of the near-surface
layers of sunspots by including available observational constraints
and semi-empirical models of the umbra and penumbra.
Applying this type of model to the sunspot of AR 9787, we showed that
the observed helioseismic signature of the sunspot model is reasonably well captured.
Our approach was to compare numerical simulations of wave propagation through the model sunspot
 and the observed SOHO/MDI cross-covariances of the Doppler velocity.

Possible improvements of the simulation include a better treatment of wave attenuation (esp. in plage),
improved initial conditions, and inclusion of the moat flow. It should also be noted that improved observations
of the surface vector magnetic field by SDO/HMI will help tune the sunspot models.

The dominant influence of the sunspot's surface layers on helioseismic waves means that it is necessary to model it
very well in order to extract information from the much weaker signature of the sunspot's
subsurface structure. The sunspot model used here accounts for most of the
helioseismic signal, although there are substantial differences that remain to be explored.
In any case, the model provides a testbed which is sufficiently similar to a real sunspot to be used for
numerous future studies in sunspot seismology.

Finally, we remark that our numerical simulations of linear waves could potentially be used to interpret other helioseismic observations than the cross-covariance (cf. helioseismic holography or Fourier-Hankel analysis), as well as other magnetic phenomena \citep[e.g., small magnetic flux tubes, see][]{Duvall2006}.

\section*{Acknowledgments}
This work is supported by the European Research Council under the European Community's Seventh Framework Programme/ERC grant agree\-ment \#210949, "Seismic Imaging of the Solar Interior", to PI L. Gizon (contribution towards Milestones 3 and 4). SOHO is a project of international collaboration between ESA and NASA.

\bibliographystyle{spr-mp-sola} 
\bibliography{helio} 


\end{document}